# The Surviving Fragment of the Zīj-i Malikshāhī: A Preliminary Edition, Translation, and Study of the 100-Star Catalogue in BnF Arabe 5968

**Rizoi Bakhromzod**[1,2*]

[1]S. U. Umarov Physical-Technical Institute, National Academy of Sciences of Tajikistan, 299/1 Aini Street, Dushanbe 734063, Republic of Tajikistan

[2]Institute of Astrophysics, National Academy of Sciences of Tajikistan 299/5 Aini Street, Dushanbe 734063, Republic of Tajikistan

*e-mail: rizo@physics.msu.ru, bakhromzod@gmail.com

**Abstract.** The astronomical programme sponsored by the Saljuq Sultan Malikshāh I in the late eleventh century occupies an important place in the history of Islamic astronomy. ʿUmar Khayyām was a principal participant in this programme, which produced the calendrical reform conventionally known as the Jalālī or Malikī calendar and appears also to have resulted in a broader set of astronomical tables known as the *Zīj-i Malikshāhī*. The complete zīj has not survived, or has not yet been securely identified. A 100-star catalogue traditionally associated with the Malikshāhī astronomical tables is preserved in the anonymous Arabic compendium *Dustūr al-munajjimīn*, in the unique manuscript BnF Arabe 5968. It consists of a catalogue of 100 fixed stars for an epoch associated with the Malikī calendrical system. The present study distinguishes this catalogue from both the Jalālī calendar and the lost or unidentified complete *Zīj-i Malikshāhī*, reconstructs its historical and manuscript context, and examines the provenance of BnF Arabe 5968 from its medieval Islamic circulation through Damascus, Bursa and Ottoman Istanbul to the collection of Charles Schefer and ultimately the Bibliothèque nationale de France. It also presents a complete transcription of the Arabic text of the catalogue from the three surviving manuscript sides, BnF Arabe 5968, fols. 224r–225r, together with a new English translation made directly from that Arabic text. B. A. Rosenfeld's Russian translation of 1961 is used throughout not as an intermediary but as a control: every entry, coordinate, magnitude and temperament abbreviation has been collated against it, and the divergences are recorded entry by entry. Rosenfeld and A. P. Yushkevich observed that 96 of the 100 stellar longitudes correspond to Ptolemaic values increased by approximately 14°26′, while 87 latitudes agree with Ptolemy. Direct manuscript collation further makes it possible to distinguish Arabic خ (Mars) from ه (Venus), both represented by Cyrillic х in Rosenfeld's edition, and thereby to recover the planetary temperaments assigned to the stars. These correspondences suggest systematic updating of inherited Ptolemaic data rather than 100 wholly independent observations, illustrating the combined roles of observation, computation, transmission and adaptation in medieval Islamic astronomy.

## 1. Introduction

ʿUmar Khayyām (1048–1131) is known internationally above all through the literary reception of the *Rubāʿiyyāt*, yet in the intellectual environment of the eleventh and early twelfth centuries he was an accomplished mathematician, astronomer and philosopher. His mathematical writings on algebra, Euclidean theory and ratios are among his best-documented scholarly achievements, while his astronomical reputation is closely associated with the calendrical reform undertaken under the Saljuq Sultan Malikshāh I. The early sources are less explicit about the precise distribution of responsibility within that programme than later popular accounts sometimes imply; consequently, Khayyām's role should be treated as that of a leading or principal participant in a collective astronomical enterprise rather than as the sole author of every astronomical product subsequently associated with Malikshāh [1].

A *zīj* was a handbook of mathematical astronomy, usually containing collections of numerical tables accompanied by instructions, parameters and explanatory material for solving astronomical problems. Depending upon the work, a zīj might contain tables for chronology, the motions of the Sun, Moon and planets, eclipses, trigonometric functions, geographical coordinates, rising times, stellar positions and astrological calculations. Such works were among the principal computational instruments of medieval Islamic astronomy. They were used in calendrical reckoning, timekeeping, planetary prediction and, in many contexts, astrology [2].

This distinction is essential for interpreting Khayyām's astronomical legacy. The **Jalālī calendar**, the broader **Malikshāhī astronomical programme**, the work known bibliographically as the ***Zīj-i Malikshāhī***, and the **surviving catalogue of 100 stars** are historically related but are not identical. The Jalālī calendar was a calendrical system; a zīj would ordinarily have been a much broader computational work; and the 100-star catalogue preserved in *Dustūr al-munajjimīn* represents a surviving table that Rosenfeld and Yushkevich associated with the wider Malikshāhī astronomical project; this association does not by itself establish Khayyām's personal authorship of every entry.

The incompleteness of the evidence also imposes a methodological restraint. The expression "Khayyām's *Zīj-i Malikshāhī*" is convenient in discussing the tradition, but it must not be taken to imply that every table within the original work was personally calculated or written by Khayyām. Nor should the surviving 100-star table be presented as the complete zīj. The more accurate designation used throughout the present study is **the surviving fragment of the *Zīj-i Malikshāhī*** or **the Malikshāhī star catalogue preserved in *Dustūr al-munajjimīn***.

The study addresses six related questions: what can be established concerning the creation and probable scope of the *Zīj-i Malikshāhī*; how it relates to the Jalālī calendar; how its surviving stellar catalogue entered *Dustūr al-munajjimīn*; what can be reconstructed about the transmission of BnF Arabe 5968; whether the stellar coordinates reflect new observations or the recalculation of Ptolemaic data; and what methodological limitations remain when a translation made directly from the Arabic witness is systematically collated with Rosenfeld's Russian edition.

## 2. The historical background of the Malikshāhī astronomical programme

The calendrical reform emerged from practical as well as theoretical problems. Lunar Islamic chronology did not remain aligned with the seasons, while the non-intercalated Persian "vague year" also shifted relative to the tropical year. These discrepancies could create difficulties in taxation and administration because fiscal practices were tied to agricultural seasons. According to the reconstruction presented by Karamati, Sultan Malikshāh I and his vizier Niẓām al-Mulk therefore assembled prominent astronomers to observe the Sun, determine the vernal equinox and establish a more seasonally stable reckoning [1].

The programme is generally associated with 467 AH/1074–1075 CE. Medieval testimony analysed by modern historians places the principal work at Isfahan. Khayyām was among the astronomers involved; sources also associate Abū al-Muẓaffar al-Isfizārī and Maymūn ibn al-Najīb al-Wāsiṭī with the enterprise, although lists of participants vary and some names repeated in later literature are less securely attested than others [1].

The observational activity should not be imagined as a single event culminating immediately in a finished calendar. Establishing a reliable solar calendar required determining the tropical year, observing the Sun near the equinox, coordinating solar motion with existing chronological systems, and devising rules capable of maintaining the beginning of the civil year near the vernal equinox. The institutional setting traditionally described as the "Isfahan observatory" was therefore part of a larger state-supported computational and observational programme.

Johannes Thomann's reassessment of the institution of the Jalālī calendar emphasises that the reform of 1079 did not simply erase older Persian calendrical practice. Rather, the new and older systems coexisted, which is important for understanding why astronomical sources could express the same epoch through several chronological eras [3].

The surviving Malikshāhī star catalogue exemplifies precisely this multi-calendar culture: its heading correlates the Malikī epoch with a "Rūmī" date and a Yazdgerd date. The table should therefore be understood not in isolation but as one element of an astronomical environment in which observational astronomy, chronological computation and administrative time-reckoning were closely connected [8].

The death of Niẓām al-Mulk and, shortly afterwards, Malikshāh in 1092 fundamentally altered the political setting that had sustained the programme. Although individual astronomers continued their scholarly activities, the institutional enterprise sponsored by Malikshāh did not persist in its earlier form. This helps to explain why the original structure and extent of the *Zīj-i Malikshāhī* are now difficult to reconstruct.

## 3. Nowruz, the Jalālī calendar, and the zodiac

The reform of the calendar was closely connected with the problem of fixing the beginning of the Persian year in relation to the vernal equinox. Bakhromzod, the author of the present article, has previously drawn attention not only to the historical circumstances of the reform but also to a modern conceptual problem that frequently distorts discussions of Khayyām: the failure to distinguish the equinoctial point, a zodiacal **sign**, and a physical astronomical **constellation** [4].

The tropical zodiac divides the ecliptic into twelve equal sectors of 30°. Its first sector, the sign Aries, begins by definition at the vernal equinoctial point. Physical constellations, by contrast, are unequal regions formed by stellar patterns and, in modern astronomy, formally delimited areas of the celestial sphere. Consequently, the statement that the Sun enters the "sign of Aries" at the vernal equinox is a statement about a conventional tropical coordinate system; it does not mean that the Sun must simultaneously enter the modern astronomical constellation Aries.

This distinction becomes particularly important because of axial precession. The orientation of the Earth's rotational axis changes slowly, causing the equinoctial points to move westward relative to the background stars. The vernal equinox therefore no longer lies among the stars of the physical constellation Aries. The present author has also argued that modern descriptions of Nowruz as the moment when the Sun "enters the constellation Aries" conflate two different systems [4]. Nowruz, in the astronomical framework relevant to the Jalālī reform, is fundamentally tied to the vernal equinox, not to entry into the physical stellar constellation Aries.

This distinction also clarifies the relationship between the calendrical reform and the Malikshāhī stellar catalogue. A solar calendar can define its beginning by a tropical phenomenon—the Sun's passage through the equinoctial point—while a star catalogue records coordinates of physical stars relative to the ecliptic. Precession affects the longitude of the stars when coordinates are referred to a moving equinoctial origin. Thus, the broader Malikshāhī astronomical milieu brought together calendrical concerns tied to the equinox and stellar computations in which longitudes had to be expressed for a specified epoch.

The present author further discussed the historical arrangement of the Jalālī months and the relation between the Sun's apparent annual motion and calendrical divisions [4]. These discussions are useful for interpreting the historical reform, but they should not lead to the assumption that the Jalālī calendar itself constituted the *Zīj-i Malikshāhī*. Rather, calendrical reform was one major output of the broader Malikshāhī enterprise.

## 4. The compilation and probable contents of the *Zīj-i Malikshāhī*

The title *Zīj-i Malikshāhī* may be translated as **"The Malikshāhī Astronomical Tables"** or, more explicitly, **"The Astronomical Tables Dedicated to Sultan Malikshāh."** The title places the work within the common Islamic tradition of naming zījes after a patron, dynasty, locality or compiler.

What the original *Zīj-i Malikshāhī* contained cannot now be reconstructed with certainty. A normal zīj of this period could include chronological tables, trigonometry, solar and lunar parameters, planetary motions, eclipse procedures, geographical data and stellar coordinates. The astronomical enterprise at Isfahan gives good reason to expect a work broader than the surviving star catalogue. Later bibliographical tradition knew a work by the Malikshāhī title, and Rosenfeld and Yushkevich treated the 100-star catalogue as material excerpted from it [6]. Their attribution is important evidence for the historiography of the table, but it should be distinguished from direct evidence of Khayyām's personal authorship.

It is therefore methodologically preferable to distinguish three levels of evidence. First, there is secure evidence for the Malikshāh-sponsored astronomical and calendrical programme. Secondly, later sources refer to a *Zīj-i Malikshāhī*. Thirdly, one identifiable table associated with that work survives in a later compilation. The existence of the third does not justify reconstructing the entire contents of the second.

## 5. What survives of the *Zīj-i Malikshāhī*?

The most concrete surviving evidence is a table of **100 fixed stars** incorporated into *Dustūr al-munajjimīn*. In BnF Arabe 5968 the catalogue occupies fols. 224r–225r (224a–225a in the alternative a/b foliation): entries 1–34 are on fol. 224r, entries 35–68 on fol. 224v, and entries 69–100 on fol. 225r [5]. The three manuscript sides examined for the present study show a carefully ruled astronomical table in red and black ink: the star descriptions occupy the broad central column, while the numerical and classificatory data are arranged in narrow vertical columns; constellation headings are written in red.

Rosenfeld's Russian edition renders the heading as positions of the fixed stars at the beginning of the first Malikī intercalary year, correlated with year 1490 of the "Rūmī" reckoning and year 448 of the Yazdgerd era [6]. His commentary interprets the latter as 1079/80 CE and the "Rūmī" era in this context as the Seleucid or "Alexander" reckoning. The Malikī era is associated there with 1079.

The table contains substantially more than positions. For each star it may give a descriptive location within the traditional figure of a constellation, the zodiacal sector, longitude, latitude, north or south latitude, magnitude, an abbreviated "temperament", and a favourable or unfavourable astrological effect. Astronomy and astrology thus occur within a single practical tabular framework, as was common in the intellectual environment in which the catalogue was copied.

The numerical zodiac signs run from 0 to 11. The sequence of stars makes their interpretation clear: 0 = Aries, 1 = Taurus, 2 = Gemini, 3 = Cancer, 4 = Leo, 5 = Virgo, 6 = Libra, 7 = Scorpio, 8 = Sagittarius, 9 = Capricorn, 10 = Aquarius, and 11 = Pisces. The numerical sign is followed by degrees and minutes within that 30° sector.

## 6. *Dustūr al-munajjimīn* and the unique manuscript

*Dustūr al-munajjimīn* is an anonymous Arabic astronomical, chronological and astrological compilation. It is **not** a work by ʿUmar Khayyām. Its importance for the present study derives from the fact that its compiler incorporated earlier materials, including the Malikshāhī stellar table.

The work reports completion at Alamut in 506 AH/1113 CE. The 2019 facsimile edition, prepared by Akbar Irani with introductory material by M. Qazvini and S. J. Badakhchani and a foreword by Farhad Daftary, describes it as a unique source associated with the early Nizārī Ismaili intellectual environment [7]. The Institute of Ismaili Studies also stresses that it is unknown whether the manuscript was among the books reportedly saved by ʿAṭā-Malik Juwaynī when Alamut fell to the Mongols in 654/1256; this is a possibility, not documented provenance [7].

The distinction between the date of composition of the work and the date of the surviving physical copy is crucial. Earlier scholarship sometimes treated the surviving BnF volume as if it were itself an early twelfth-century Alamut manuscript. Boris Liebrenz, however, states that on palaeographical grounds the extant copy is clearly not a manuscript of the fifth to early sixth century AH [10]. The BnF/Biblissima description identifies it as a fourteenth-century naskh copy comprising 346 folios and measuring approximately 23 × 16 cm [11].

This changes the historical question. The text may preserve material compiled at Alamut in 1113, but the codex presently in Paris is a later witness to that text. It must therefore be studied as an artefact with its own subsequent history.

## 7. The history and provenance of BnF Arabe 5968

Liebrenz's reassessment is especially important because it replaces a simple assumed east-to-west path with a reconstruction grounded in ownership inscriptions, reader's notes, palaeography and biographical evidence [10]. His analysis shows both what can be recovered and how uncertain several stages remain.

The earliest dated annotation identified by Liebrenz occurs on fol. 255v and is dated Rajab 773 AH. Although it is a historical gloss rather than a formal ownership statement, it provides a useful *terminus ante quem*: the surviving codex must already have existed by that date. This agrees broadly with the BnF catalogue's fourteenth-century dating.

An undated ownership statement on the title page can be read, according to Liebrenz, as that of **Muḥammad b. Yūsuf b. al-Ḥawrānī al-Ḥanbalī**. His nisba and legal affiliation suggest a Syrian context, possibly the Ḥanbalī community of Damascus, although this localisation remains inferential rather than certain.

A much more securely identifiable owner was **Jalāl al-Dīn Muḥammad b. Muḥammad al-Ramlī** (d. 1000/1591–92). Despite the nisba referring ultimately to Ramla, he lived in Damascus and served as a *muwaqqit*, a specialist responsible for astronomical timekeeping, at the Umayyad Mosque. His possession of an astronomical compendium is therefore particularly intelligible in relation to his professional activity.

Another ownership note belongs to **Aḥmad b. Muḥammad al-Wānī**. Liebrenz emphasises that the nisba al-Wānī refers to Van, rather than being an erroneous form of al-Shīrwānī; it does not, however, prove that Aḥmad himself resided in Van.

The manuscript next appears in a note by **Murtaḍā b. Ḥasan**, who states that he obtained possession when the volume was brought to Bursa. A damaged word may possibly indicate an auction, but Liebrenz explicitly treats that reconstruction as speculative.

At another stage the volume entered the celebrated library of **Abū Bakr b. Rustam b. Aḥmad b. Maḥmūd al-Shīrwānī** (d. 1135/1723), an Ottoman administrator and major bibliophile in Istanbul. Liebrenz identifies this ownership with a well-documented seventeenth- and early eighteenth-century Istanbul collection rather than assuming from the nisba that the manuscript was then physically located in Shirvan.

More than a century remains undocumented between al-Shīrwānī's death and the manuscript's acquisition by the French Orientalist and diplomat **Charles Schefer (1820–1898)** sometime in the nineteenth century. Liebrenz explicitly notes this lacuna. The Bibliothèque nationale subsequently acquired Schefer's collection in **1899** [11].

**Table 1. Reconstructed provenance of BnF Arabe 5968**

| Approximate date | Owner or institution | Place | Evidence | Source | Degree of certainty |
|---|---|---|---|---|---|
| 506 AH/1113 CE | Composition of *Dustūr al-munajjimīn* | Alamut | Internal textual tradition concerning completion | [7] | Documented for composition of the work, not the surviving codex |
| Before Rajab 773 AH | Surviving physical manuscript already in existence | Unknown | Dated marginal gloss, fol. 255v | [10] | Documented |
| Possibly fourteenth century | Muḥammad b. Yūsuf b. al-Ḥawrānī al-Ḥanbalī | Probably Syria/Damascus | Ownership inscription; nisba and legal affiliation | [10] | Probable |
| Before 1000 AH/1591–92 | Jalāl al-Dīn Muḥammad al-Ramlī | Damascus | Ownership inscription and secure biographical identification | [10] | Documented |

| Undated | Aḥmad b. Muḥammad al-Wānī | Uncertain | Ownership inscription | [10] | Documented owner; location uncertain |
|---|---|---|---|---|---|
| Undated | Murtaḍā b. Ḥasan | Bursa | Ownership note referring to arrival of book in Bursa | [10] | Probable; auction hypothesis uncertain |
| Before 1135 AH/1723 | Abū Bakr al-Shīrwānī | Istanbul | Ownership inscription; identifiable private library | [10] | Documented |
| Mid-nineteenth century | Charles Schefer | Ottoman/French collecting networks | Schefer provenance; precise acquisition circumstances unknown | [10, 11] | Documented ownership; acquisition route uncertain |
| 1899 onward | Bibliothèque nationale de France | Paris | Institutional acquisition of Schefer collection | [11] | Documented |

A possible connection between the textual tradition of *Dustūr al-munajjimīn* and manuscripts reportedly preserved by Juwaynī after the fall of Alamut has been proposed, but no ownership or codicological evidence directly connects the extant BnF Arabe 5968 with that event.

On the basis of the surviving ownership and reading notes, a cautious reconstruction of the manuscript's recoverable trajectory is Syria → Bursa → Istanbul → Paris, although the precise order and chronology of some intermediate stages remain uncertain. The history of the manuscript thus illustrates the mobility of scientific books across private libraries, mosque-centred scholarly communities, Ottoman book markets and nineteenth-century Orientalist collecting.

## 8. Astronomical characteristics of the 100-star catalogue

Rosenfeld and Yushkevich's most significant astronomical observation concerns the relationship between the Malikshāhī positions and Ptolemy's *Almagest* [12]. According to their comparison, 96 of the 100 stellar longitudes are greater than Ptolemy's corresponding values by approximately 14°26′, while 87 of the 100 stellar latitudes coincide with the Ptolemaic latitudes [6]. The Russian commentary explicitly gives both figures.

This interpretation is consistent with the wider Islamic astronomical tradition. In their study of the star table accompanying al-Fārisī's Zīj al-Muẓaffarī, Kunitzsch and Langermann define the 'temperaments' of fixed stars as the planet or planets associated with them and note that different traditions concerning these planetary assignments circulated in medieval astronomical sources [14].

This pattern is highly informative. Under a simple precessional updating of an inherited ecliptic catalogue, stellar latitudes may remain nearly unchanged, whereas longitudes referred to the moving equinoctial origin undergo a systematic shift; proper motion and smaller coordinate effects can nevertheless introduce additional differences. A catalogue produced for a later epoch could therefore be generated by retaining much of the inherited latitude data while adding a precessional increment to older longitudes. The high proportion of exact or near-exact correspondences strongly suggests that the catalogue was not constructed from 100 wholly independent fresh determinations of both longitude and latitude.

The conclusion should not, however, be that the catalogue is scientifically unimportant because it is "derived". Updating earlier tables was itself a fundamental astronomical operation. Medieval astronomers worked within a cumulative mathematical tradition: observation, correction, recalculation, interpolation and the adaptation of older parameters for a new epoch were all legitimate scientific practices. The authority of Ptolemy's stellar catalogue remained considerable, yet inherited data were not necessarily reproduced unchanged.

Comparable procedures are documented in other Islamic zījes. Kunitzsch and Langermann show, for example, that derivative versions of al-Fārisī's stellar table preserved essentially the same star list while systematically increasing the stellar longitudes for later epochs [14]. Such evidence demonstrates that the epochal updating of inherited stellar coordinates was a recognised feature of the zīj tradition.

A shift of 14°26′ also has the expected order of magnitude for the accumulated effect attributed to precession between the epoch of the Ptolemaic catalogue and the late eleventh century. The catalogue therefore illustrates in numerical form the same astronomical phenomenon that makes it necessary to distinguish the tropical equinox from fixed constellations in calendrical discussions.

Differences among the remaining stars may have several origins: fresh observation; alternative source catalogues; computational rounding; scribal error; corruption during transmission; or imperfect correspondence between medieval descriptive identifications and modern stellar identifications. Without a new critical edition comparing every Arabic coordinate against Ptolemy and other Islamic catalogues, the precise cause of each deviation cannot be assigned. For the broader debate concerning the formation and observational basis of Ptolemy's star catalogue, see Evans [13].

The catalogue also preserves Ptolemaic-style stellar magnitudes. Rosenfeld and Yushkevich explain that the signs represented in their Russian edition by **б** and **м** correspond to Arabic *kāf* and *ṣād*, initials indicating *kabīr* ("greater") and *ṣaghīr* ("smaller") relative to the stated magnitude [6]. Rosenfeld and Yushkevich considered the exact meanings of the *mizāj* abbreviations not fully clear. Direct collation with BnF Arabe 5968, however, shows that these signs encode the planetary temperament assigned to each star. The abbreviations are formed from letters associated with the Arabic planet names; importantly, Rosenfeld's Cyrillic х may represent either Arabic خ (Mars) or ہ (Venus), a distinction recoverable only from the manuscript.

## 9. Editorial principles and translation methodology

The English translation presented below is a **new translation made directly from the Arabic** of BnF Arabe 5968, fols. 224r–225r. All three sides were read in the digital facsimile published by the Bibliothèque nationale de France on Gallica [5]. For each of the 100 entries the following elements were transcribed independently from the manuscript: the descriptive designation of the star in the broad central column; the red marginal rubric naming the constellation; the three subcolumns of longitude (*al-burūj*, *al-daraj*, *al-daqāʾiq*) and the two subcolumns of latitude, all written in abjad numerals; the direction column (*al-jihāt*); the magnitude column (*al-aqdār*); the temperament column (*al-ṭabāʾiʿ*); and the effects column (*al-afʿāl*).

B. A. Rosenfeld's Russian translation, published in *Omar Khayyam: Traktaty* (Moscow, 1961), is used here **as a control rather than as an intermediary** [6]. Once the Arabic text had been established, every entry was collated against the 1961 edition; agreements are passed over in silence, while each divergence in wording, coordinate, magnitude or temperament is recorded in a dedicated column of the translation table and, where it is astronomically significant, discussed in the apparatus. This procedure differs from that of an indirect translation, and it makes the two witnesses — the Arabic manuscript and the Russian edition — independently verifiable by the reader.

The Arabic text is presented first, in tabular form reproducing the layout of the manuscript (§10.3); the English translation follows (§10.4). The following conventions are used:

1. Angle brackets ⟨…⟩ mark readings that are uncertain because the ink has faded, the parchment is stained, or the word is obscured by the gutter; a question mark inside the brackets marks a conjectural word.
2. Square brackets indicate supplied editorial text.
3. Numbers are given as they stand in the manuscript, in abjad, with the arabic-numeral equivalent in parentheses. Zodiacal signs are counted from zero, so that ٠ = Aries and يا = Pisces; the sign is followed by degrees and minutes within the 30° sector.
4. In the magnitude column ك (*kabīr*, "greater") and ص (*ṣaghīr*, "smaller") qualify the preceding numeral, corresponding to Ptolemy's μείζων and ἐλάσσων; سحابي denotes a nebulous object. These are the signs that Rosenfeld renders by Cyrillic **б** and **м**.
5. In the temperament column each letter is the final letter of the Arabic name of a planet (see the key in §10.1). Rosenfeld's Cyrillic **х** renders two distinct Arabic letters, خ (Mars) and ہ (Venus); the distinction is recoverable only from the manuscript and is restored throughout.
6. In the effects column the manuscript groups most entries under a brace; individual entries marked قاطع are rendered "unfavourable". The reading of the brace word itself is not yet secure, and the column is therefore reported conservatively.
7. Medieval coordinates are retained and are not replaced by modern astronomical positions; "favourable" and "unfavourable" are historical astrological categories, not modern astronomical assessments.

The present edition remains a working edition rather than a full diplomatic transcription: the numerical columns have been read throughout, but a definitive text would require collation with any further witnesses of *Dustūr al-munajjimīn* and a systematic comparison of every coordinate with the *Almagest*.

## 10. The Arabic text and a new English translation of the surviving fragment

### 10.1 Rosenfeld's title and the manuscript heading

**Rosenfeld's editorial title: THE MALIKSHĀHĪ ASTRONOMICAL TABLES**

**Positions of the fixed stars at the beginning of [the first] intercalary year of the Malikī [reckoning], that is, year 1490 of the Rūmī [Seleucid] era and year 448 of Yazdgerd.**

The following key is derived from direct manuscript comparison and from the established medieval convention of expressing stellar temperaments through associated planets:

| Arabic abbreviation | Arabic planet name | Planet | Rosenfeld |
|---|---|---|---|
| ل | زحل | Saturn | л |
| ي | المشتري | Jupiter | и/й |
| خ | المريخ | Mars | х |
| س | الشمس | Sun | с |
| ه | الزهرة | Venus | х |
| د | عطارد | Mercury | д |
| ر | القمر | Moon | р |

### 10.2 The three surviving folios

The catalogue occupies three consecutive sides of BnF Arabe 5968. They are reproduced below from the digital facsimile of the Bibliothèque nationale de France, where the volume is available in full: https://gallica.bnf.fr/ark:/12148/btv1b525110398/f456.item. The three sides correspond to images f453, f454 and f455 of that digitisation.

Source gallica.bnf.fr / Bibliothèque nationale de France. Département des Manuscrits. Arabe 5968

***Fig. 1.*** *BnF, Arabe 5968, fol. 224r (Gallica image f453): heading of the table and entries 1–34. Source: gallica.bnf.fr / Bibliothèque nationale de France.*

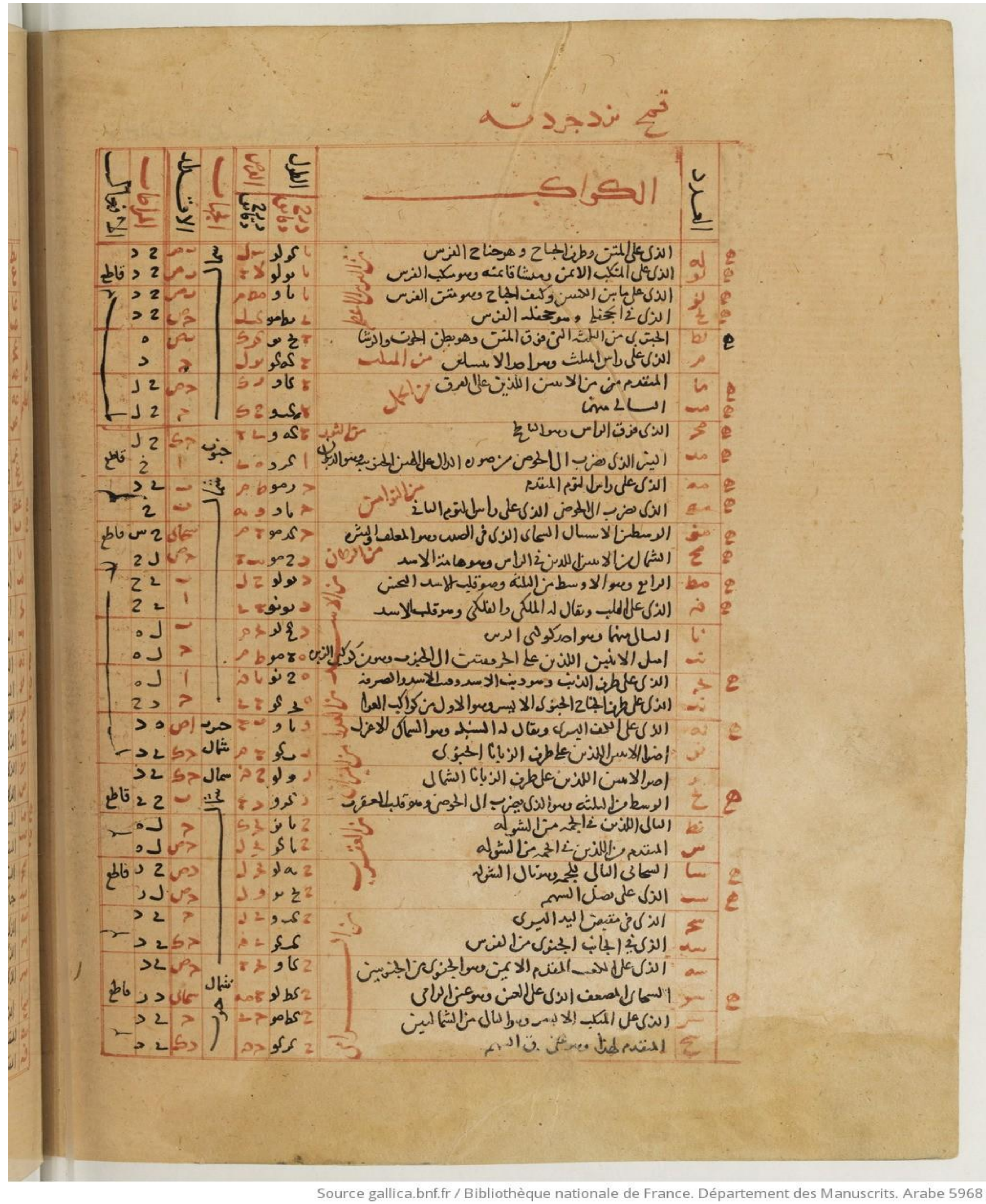

***Fig. 2.*** *BnF, Arabe 5968, fol. 224v (Gallica image f454): entries 35–68; Source: gallica.bnf.fr / Bibliothèque nationale de France.*

***Fig. 3.*** *BnF, Arabe 5968, fol. 225r (Gallica image f455): entries 69–100, the end of the catalogue. Source: gallica.bnf.fr / Bibliothèque nationale de France.*

The layout is identical on all three sides. Reading from the right: the entry number in abjad, written in red; the descriptive designation of the star in black, with the constellation rubric added in red in the margin; the three subcolumns of longitude; the two subcolumns of latitude; the direction column, in which a single word (شمال or جنوب) is written vertically and covers a group of entries under a brace; the magnitude column; the temperament column; and the effects column, likewise braced.

## 10.3 The Arabic text of the catalogue

The table below reproduces the text of the three folios. Numbers are given in abjad with the arabic-numeral equivalent in parentheses; the columns follow the manuscript order.

**Heading (fol. 224r):** مواضع الثوابت لأوّل سنة الكبيسة الملكيّة وهي سنة ⟨...⟩ روميّة ⟨...⟩ تمح يزدجرديّه

**Column headings:** العدد ‖اسماء الكواكب ومواضعها من الصور ‖الطول :البروج /الدرج /الدقايق ‖العرض :الدرج /الدقايق ‖الجهات ‖الاقدار ‖الطبايع ‖الافعال

| № | الكواكب ومواضعها من الصور | الصورة | الطول )البروج/الدرج/الدقايق( | العرض )الدرج/الدقايق( | الجهات | الاقدار | الطبايع | الافعال |
|---|---|---|---|---|---|---|---|---|
| ا — ١ | انور الفرقدين | من الدب الاصغر | د/ا/نو — ٤/١/٥٦ | عب/ن — ٧٢/٥٠ | شمال | ب | له | — |
| ب — ٢ | الآخر من الفرقدين | — | د/ي/لو — ٤/١٠/٣٦ | عد/ن — ٧٤/٥٠ | شمال | ج | هد | — |
| ج — ٣ | الذي على طرف الذنب وهو الجدي | — | ب/يد/لو — ٢/١٤/٣٦ | سو/٠ — ٦٦/٠ | شمال | ج | ل | — |
| د — ٤ | الذي على الظهر من التي في ذي الاربعة بعد الاصلاح | من الدب الاكبر | د/يد/و — ٤/١٤/٦ | مط/٠ — ٤٩/٠ | شمال | ب | خ | — |
| ه — ٥ | الذي على المراقّ منها | — | د/و/لو — ٤/٦/٣٦ | مه/ل — ٤٥/٣٠ | شمال | جك | خ | — |
| و — ٦ | الذي على معقد الذنب منها | — | د/يز/لو — ٤/١٧/٣٦ | نا/٠ — ٥١/٠ | شمال | جص | خ | — |
| ز — ٧ | الباقية منها وهو على الفخذ اليسرى المؤخّر | — | د/يز/كو — ٤/١٧/٢٦ | مو/ل — ٤٦/٣٠ | شمال | جص | خ | — |
| ح — ٨ | الاول من الثلثة التي تلي معقد الذنب وهو الجون | — | د/كو/لو — ٤/٢٦/٣٦ | نج/ل — ٥٣/٣٠ | شمال | ب | هد | — |
| ط — ٩ | الاوسط منها وهو العناق | — | ه/ب/كو — ٥/٢/٢٦ | نه/م — ٥٥/٤٠ | شمال | ب | هد | — |
| ي — ١٠ | الثالث وهو الذي على طرف الذنب وهو القائد | — | ه/يد/يو — ٥/١٤/١٦ | ند/٠ — ٥٤/٠ | شمال | ب | هد | — |
| يا — ١١ | احدى كوكبي )الدسن؟( وهو المائل الى الشمال | من التنين | ه/كب/مو — ٥/٢٢/٤٦ | فد/ن — ٨٤/٥٠ | شمال | ج | لخ | — |
| يب — ١٢ | )الشمالي( من الاثنين اللذين ممّا يلي المغرب الى الشمال | — | ه/كد/و — ٥/٢٤/٦ | عح/٠ — ٧٨/٠ | شمال | ج | لخ | — |
| يج — ١٣ | اللامس من فوق المنكب الايمن وهو احد كوكبي الغرب | من قيفاوس | ٠/ا/و — ٠/١/٦ | سط/٠ — ٦٩/٠ | شمال | ج | له | — |
| يد — ١٤ | الذي فيما بين فخذي العوّا وهو السماك الرامح | من العوّا | و/يا/كو — ٦/١١/٢٦ | لا/ل — ٣١/٣٠ | شمال | ا | يخ | — |
| يه — ١٥ | النيّر )الاجمل؟( وهو )نيّر( الفكّة | من الاكليل | و/كط/و — ٦/٢٩/٦ | مد/ل — ٤٤/٣٠ | شمال | ب | هد | — |
| يو — ١٦ | الذي على الرأس وهو كلب الراعي | من الجاثي على ركبتيه | ح/ب/و — ٨/٢/٦ | لز/ل — ٣٧/٣٠ | شمال | جص | هد | — |
| يز — ١٧ | الذي على )الحرب؟( وهو النسر الواقع ويقال له اللورا | من الصنج | ط/ا/مو — ٩/١/٤٦ | سب/٠ — ٦٢/٠ | شمال | ا | هد | — |
| يح — ١٨ | الذي على الفم وهو منقار الدجاجة | من الدجاجة | ط/يح/نو — ٩/١٨/٥٦ | مط/ك — ٤٩/٢٠ | شمال | جص | هد | — |
| يط — ١٩ | النيّر الذي في الذنب وهو الردف | — | ي/كج/لو — ١٠/٢٣/٣٦ | س/٠ — ٦٠/٠ | شمال | ب | هد | — |
| ك — ٢٠ | الذي في وسط المسند وهو على سنام الناقة وهو كفّ الخضيب | من ذات الكرسي | ٠/كب/يو — ٠/٢٢/١٦ | نا/م — ٥١/٤٠ | شمال | ج | له | — |
| كا — ٢١ | الشمالي السحابي الذي على طرف اليد اليمنى وهو معصم الثريا | — | ا/يا/و — ١/١١/٦ | م/ل — ٤٠/٣٠ | شمال | سحابي | خد | — |
| كب — ٢٢ | النيّر الذي في الجنب الايمن وهو مرفق الثريا | — | ا/يط/يو — ١/١٩/١٦ | ل/٠ — ٣٠/٠ | شمال | ب | خد | — |
| كج — ٢٣ | النيّر الذي في رأس الغول | — | ا/يد/و — ١/١٤/٦ | كج/٠ — ٢٣/٠ | شمال | بص | لد | — |
| كد — ٢٤ | الذي على المنكب الايسر وهو العيوق | من ممسك الاعنّة | ب/ط/كو — ٢/٩/٢٦ | كب/ل — ٢٢/٣٠ | شمال | ا | خد | — |
| كه — ٢٥ | الذي على المنكب الايمن وهو احد الاعلام وما يلي النجم | — | ب/يز/يو — ٢/١٧/١٦ | ك/٠ — ٢٠/٠ | شمال | ب | خد | — |
| كو — ٢٦ | الذي على الكعب الايمن وهو المشترك له وللقرن الشمالي من الثور | — | ب/ي/و — ٢/١٠/٦ | ه/٠ — ٥/٠ | شمال | ب | خد | — |
| كز — ٢٧ | الذي على الرأس وهو الراعي | من الحوّاء والحيّة | ح/ط/يو — ٨/٩/١٦ | لو/٠ — ٣٦/٠ | شمال | ج | له | — |
| كح — ٢٨ | المتقدّم من الاثنين اللذين على المنكب الايمن وهو كليب الراعي | — | ح/يب/كو — ٨/١٢/٢٦ | كز/يه — ٢٧/١٥ | شمال | جص | له | — |
| كط — ٢٩ | الاوسط من الثلثة التي على عنق الحيّة من الصفّ الشآمي وهو عنق الحيّة | — | ز/ح/مو — ٧/٨/٤٦ | كه/ك — ٢٥/٢٠ | شمال | ج | له | — |
| ل — ٣٠ | المفرد الذي على النصل | من السهم | ط/كد/لو — ٩/٢٤/٣٦ | لط/ك — ٣٩/٢٠ | شمال | ج | خه | — |
| لا — ٣١ | النيّر الذي فيما بين المنكبين وهو النسر الطائر | من العقاب | ط/يح/يو — ٩/١٨/١٦ | كط/ي — ٢٩/١٠ | شمال | بك | خي | — |
| لب — ٣٢ | المتقدّم من الثلثة التي في الذنب وهو ذنب الدلفين | من الدلفين | ي/ب/و — ١٠/٢/٦ | كط/ي — ٢٩/١٠ | شمال | دك | خه | — |
| لج — ٣٣ | المتقدّم من الاثنين اللذين في الرأس | من قطعة الفرس | ي/ي/مو — ١٠/١٠/٤٦ | ك/ل — ٢٠/٣٠ | شمال | د | ل | — |
| لد ـــ ٣٤ | الذي على السرّة وهو مشترك لها ولرأس المرأة المسلسلة | من الفرس الاعظم | ٠/ب/يو ـــ ٠/٢/١٦ | كو/٠ ـــ ٢٦/٠ | شمال | بص | خد | ـــ |
| له ـــ ٣٥ | الذي على المتن وطرف الجناح وهو جناح الفرس | ـــ | يا/كو/لو ـــ ١١/٢٦/٣٦ | يب/ل ـــ ١٢/٣٠ | شمال | بص | خد | قاطع |
| لو ـــ ٣٦ | الذي على المنكب الايمن ومبدأ الساق منه وهو منكب الفرس | ـــ | يا/يو/لو ـــ ١١/١٦/٣٦ | لا/٠ ـــ ٣١/٠ | شمال | بص | خد | قاطع |
| لز ـــ ٣٧ | الذي على ما بين الكتفين وكتف الجناح وهو متن الفرس | ـــ | يا/يا/و ـــ ١١/١١/٦ | يط/م ـــ ١٩/٤٠ | شمال | بص | خد | ـــ |
| لح ـــ ٣٨ | الذي في الجحفلة وهو جحفلة الفرس | ـــ | ي/يط/مو ـــ ١٠/١٩/٤٦ | كب/ل ـــ ٢٢/٣٠ | شمال | جص | خد | ـــ |
| لط ـــ ٣٩ | الجنوبي من الثلثة التي فوق المتن وهو بطن الحوت والرشا | ـــ | ٠/يح/يو ـــ ٠/١٨/١٦ | كو/ك ـــ ٢٦/٢٠ | شمال | بص | ه | ـــ |
| م ـــ ٤٠ | الذي على رأس المثلّث وهو احد )الاساس؟( | من المثلّث | ٠/كه/كو ـــ ٠/٢٥/٢٦ | يو/ل ـــ ١٦/٣٠ | شمال | ج | د | ـــ |
| ما ـــ ٤١ | المتقدّم من الاثنين اللذين على القرن | من الحمل | ٠/كا/و ـــ ٠/٢١/٦ | ز/ك ـــ ٧/٢٠ | شمال | جص | خل | ـــ |
| مب ـــ ٤٢ | التالي منها | ـــ | ٠/كب/و ـــ ٠/٢٢/٦ | ح/ك ـــ ٨/٢٠ | جنوب | ج | خل | ـــ |
| مج ـــ ٤٣ | الذي فوق الرأس وهو الناطح | ـــ | ٠/كه/و ـــ ٠/٢٥/٦ | ي/٠ ـــ ١٠/٠ | جنوب | جك | خل | قاطع |
| مد ـــ ٤٤ | النيّر الذي يضرب الى الحمرة المصوّر في الدال على العين الجنوبية وهو الدبران | من الثور | ا/كز/و ـــ ١/٢٧/٦ | ه/ي ـــ ٥/١٠ | جنوب | ا | خ | قاطع |
| مه ـــ ٤٥ | الذي على رأس التوأم المتقدّم | ـــ | ج/و/مو ـــ ٣/٦/٤٦ | ط/م ـــ ٩/٤٠ | شمال | ب | يد)؟( | ـــ |
| مو ـــ ٤٦ | الذي يضرب الى الحمرة الذي على رأس التوأم الثاني | من التوأمين | ج/يا/و ـــ ٣/١١/٦ | و/يه ـــ ٦/١٥ | شمال | ب | خ | ـــ |
| مز ـــ ٤٧ | الاوسط السحابي الذي في الصدر وهو المعلف والنثرة | من السرطان | ج/كد/مو ـــ ٣/٢٤/٤٦ | ٠/م ـــ ٠/٤٠ | شمال | سحابي | خس | قاطع |
| مح ـــ ٤٨ | الشمالي من الاثنين اللذين في الرأس وهو هامة الاسد | من الاسد | د/ح/مو ـــ ٤/٨/٤٦ | يب/٠ ـــ ١٢/٠ | شمال | جص | لخ | ـــ |
| مط ـــ ٤٩ | الرابع وهو الاوسط من الثلثة وهو )...( الاسد )المحسن؟( | ـــ | د/ح/لو ـــ ٤/٨/٣٦ | ح/ل ـــ ٨/٣٠ | شمال | ا | يخ | ـــ |
| ن ـــ ٥٠ | الذي على القلب ويقال له الملكي والفلكي وهو قلب الاسد | ـــ | د/يو/نو ـــ ٤/١٦/٥٦ | ٠/ي ـــ ٠/١٠ | شمال | ا | يخ | ـــ |
| نا ـــ ٥١ | التالي منها وهو احد )ركبتي( الاسد | ـــ | د/كح/لو ـــ ٤/٢٨/٣٦ | يج/م ـــ ١٣/٤٠ | شمال | ب | له | ـــ |
| نب ـــ ٥٢ | اسفل الاثنين اللذين على آخر )مؤخّرتي الحجزين؟( وهو من الزبرة | ـــ | ه/٠/مو ـــ ٥/٠/٤٦ | ط/م ـــ ٩/٤٠ | شمال | ج | له | ـــ |
| نج ـــ ٥٣ | الذي على طرف الذنب وهو ذنب الاسد وصلب الاسد والصرفة | ـــ | ه/ح/نو ـــ ٥/٨/٥٦ | يا/ن ـــ ١١/٥٠ | شمال | ا | له | ـــ |
| ند ـــ ٥٤ | الذي على طرف الجناح الجنوبي الايسر وهو الاول من كواكب العوّا | من السنبلة | ه/يج/كو ـــ ٥/١٣/٢٦ | ٠/ي ـــ ٠/١٠ | شمال | ج | دخ | ـــ |
| نه ـــ ٥٥ | الذي على الكفّ الايسر ويقال له السنبلة وهو السماك الاعزل | ـــ | و/يا/و ـــ ٦/١١/٦ | ب/٠ ـــ ٢/٠ | جنوب | اص | ده | ـــ |
| نو ـــ ٥٦ | اضوأ الاثنين اللذين على طرف الزبانى الجنوبي | من الميزان | ز/ب/كو ـــ ٧/٢/٢٦ | ٠/م ـــ ٠/٤٠ | شمال | جك | يد | ـــ |
| نز ـــ ٥٧ | اضوأ الاثنين اللذين على طرف الزبانى الشمالي | ـــ | ز/و/لو ـــ ٧/٦/٣٦ | ح/ن ـــ ٨/٥٠ | شمال | جك | يد | قاطع |
| نح ـــ ٥٨ | الاوسط من الثلثة وهو الذي يضرب الى الحمرة وهو قلب العقرب | من العقرب | ز/كز/و ـــ ٧/٢٧/٦ | د/٠ ـــ ٤/٠ | جنوب | ب | خي | قاطع |
| نط ـــ ٥٩ | التالي من اللذين في الحمة من الشولة | ـــ | ح/يا/نو ـــ ٨/١١/٥٦ | يج/ك ـــ ١٣/٢٠ | جنوب | ج | لخ)؟( | ـــ |
| س ـــ ٦٠ | المتقدّم من اللذين في الحمة من الشولة | ـــ | ح/يا/كو ـــ ٨/١١/٢٦ | يج/ل ـــ ١٣/٣٠ | جنوب | جص | له)؟( | ـــ |
| سا ـــ ٦١ | السحابي التالي للحمة وهو )نيال الشولة؟( | من القوس | ح/يه/لو ـــ ٨/١٥/٣٦ | يج/ل ـــ ١٣/٣٠ | جنوب | دص | خد | قاطع |
| سب ـــ ٦٢ | الذي على نصل السهم | ـــ | ح/يح/نو ـــ ٨/١٨/٥٦ | و/ل ـــ ٦/٣٠ | جنوب | جص | لر | قاطع |
| سج ـــ ٦٣ | الذي في مقبض اليد اليسرى | ـــ | ح/كب/و ـــ ٨/٢٢/٦ | ي/ل ـــ ١٠/٣٠ | جنوب | ج | يد | ـــ |
| سد ـــ ٦٤ | الذي في الجانب الجنوبي من القوس | ـــ | ح/كب/كو ـــ ٨/٢٢/٢٦ | ي/ن ـــ ١٠/٥٠ | جنوب | جك | يد | ـــ |
| سه ـــ ٦٥ | الذي على )لحب؟( المقدّم الايمن وهو الجنوبي من الجنوبيين | ـــ | ح/كا/و ـــ ٨/٢١/٦ | يج/٠ ـــ ١٣/٠ | جنوب | جص | يد | قاطع |
| سو ـــ ٦٦ | السحابي المصعّف الذي على العين وهو عين الرامي | ـــ | ح/كط/لو ـــ ٨/٢٩/٣٦ | ٠/مه ـــ ٠/٤٥ | شمال | سحابي | در | قاطع |
| سز ـــ ٦٧ | الذي على المنكب الايسر والتالي من الشماليين | ـــ | ح/كط/مو ـــ ٨/٢٩/٤٦ | ج/ي ـــ ٣/١٠ | جنوب | جص | يد | ـــ |
| سح ـــ ٦٨ | المتقدّم لهما وهو مشترك في )القدح؟( | ـــ | ح/كز/كو ـــ ٨/٢٧/٢٦ | ج/ن ـــ ٣/٥٠ | جنوب | دك | يد | ـــ |
| سط ـــ ٦٩ | الاوسط منها وهو على الكتف | ـــ | ط/ب/و ـــ ٩/٢/٦ | د/ل ـــ ٤/٣٠ | جنوب | دك | يد | ـــ |
| ع ـــ ٧٠ | )الثالثة( وهو تحت الابط | من الرامي | ط/٠/مو ـــ ٩/٠/٤٦ | و/مه ـــ ٦/٤٥ | جنوب | ج | يد | ـــ |
| عا ـــ ٧١ | الشمالي من الثلثة التي في القرن التالي | من الجدي | ط/كا/مو ـــ ٩/٢١/٤٦ | ز/ك ـــ ٧/٢٠ | شمال | جص | هخ | ـــ |
| عب ـــ ٧٢ | الجنوبي من الثلثة | ـــ | ط/كا/مو ـــ ٩/٢١/٤٦ | ه/٠ ـــ ٥/٠ | شمال | جص | هخ | ـــ |
| عج ـــ ٧٣ | الذي في المنكب الايسر وهو من سعد السعود | من الدالي | ي/ي/نو ـــ ١٠/١٠/٥٦ | ح/ن ـــ ٨/٥٠ | شمال | جص | لد | ـــ |
| عد ـــ ٧٤ | الذي في آخر الماء وهو على فم الحوت الجنوبي وهو الضفدع الاول | ـــ | ي/كا/كو ـــ ١٠/٢١/٢٦ | كج/٠ ـــ ٢٣/٠ | جنوب | ا | ره)؟( | ـــ |
| عه ـــ ٧٥ | الذي في فم السمكة المتقدّمة | من السمكتين | يا/و/و ـــ ١١/٦/٦ | ط/يه ـــ ٩/١٥ | شمال | د | دل | ـــ |
| عو — ٧٦ | الذي على عقد الخيطين | — | ٠/يو/نو — ٠/١٦/٥٦ | ح/ل — ٨/٣٠ | جنوب | جص | دل | — |
| عز — ٧٧ | الذي في الشعبة الشمالية من اللذين في ⟩سمكة⟨ الذنب وهو ذنب قيطس | — | يا/يح/مو — ١١/١٨/٤٦ | ط/م — ٩/٤٠ | جنوب | جص | ل | — |
| عح — ٧٨ | الذي في الشعبة الجنوبية منه من الذنب وهو الضفدع الثاني | — | يا/ك/و — ١١/٢٠/٦ | ك/ك — ٢٠/٢٠ | جنوب | جك | ل | — |
| عط — ٧٩ | ⟩السحابي⟨ الذي في رأس الجبار وهو الهقعة ورأس الجبار | — | ب/يا/كو — ٢/١١/٢٦ | يج/ن — ١٣/٥٠ | جنوب | سحابي | خد | قاطع |
| ف — ٨٠ | النيّر الذي على المنكب الايمن وهو يد الجوزا ومنكب الجوزا | — | ب/يو/كو — ٢/١٦/٢٦ | يز/٠ — ١٧/٠ | جنوب | اص | خد | قاطع |
| فا — ٨١ | الذي على المنكب الايسر وهو الناجذ والمرزم | — | ب/ح/كو — ٢/٨/٢٦ | يز/ل — ١٧/٣٠ | جنوب | ب | يل | — |
| فب — ٨٢ | المتقدّم من الثلثة التي على المنطقة | — | ب/ط/مو — ٢/٩/٤٦ | كد/ي — ٢٤/١٠ | جنوب | ب | يل | — |
| فج — ٨٣ | الاوسط منها | — | ب/يا/مو — ٢/١١/٤٦ | كد/ن — ٢٤/٥٠ | جنوب | ب | يل | — |
| فد — ٨٤ | التالي من الثلثة | — | ب/يب/لو — ٢/١٢/٣٦ | كه/م — ٢٥/٤٠ | جنوب | ب | يل | — |
| فه — ٨٥ | النيّر الذي في القدم اليسرى وهو رجل الجوزا اليسرى وراعي الجوزا | — | ب/د/يو — ٢/٤/١٦ | لا/ل — ٣١/٣٠ | جنوب | ا | يل | قاطع |
| فو — ٨٦ | النيّر الذي في آخر النهر وهو الظليم | من النهر | ٠/يد/لو — ٠/١٤/٣٦ | نج/ل — ٥٣/٣٠ | جنوب | ا | يه | — |
| فز — ٨٧ | الذي في وسط البدن | من الارنب | ب/ي/يو — ٢/١٠/١٦ | ما/ل — ٤١/٣٠ | جنوب | جص | لخ | — |
| فح — ٨٨ | الذي على الفم وهو الشعرى اليمانية والعبور وكلب الجبار | من الكلب الاكبر | ج/ب/و — ٣/٢/٦ | لط/ي — ٣٩/١٠ | جنوب | ا | يخ | — |
| فط — ٨٩ | الذي على طرف اليد المتقدّمة اليمنى وهو المرزم | — | ب/كه/كو — ٢/٢٥/٢٦ | ما/ك — ٤١/٢٠ | جنوب | ج | يخ | — |
| ص — ٩٠ | الذي في الجيد وهو المرزم | من الكلب الاصغر | ج/ط/كو — ٣/٩/٢٦ | يد/٠ — ١٤/٠ | جنوب | د | خد | — |
| صا — ٩١ | النيّر الذي على ⟩العجز⟨ وهو الشعرى الشامية والغميصاء | — | ج/يج/لو — ٣/١٣/٣٦ | يو/ي — ١٦/١٠ | جنوب | ا | خد | — |
| صب — ٩٢ | المتقدّم من الاثنين اللذين في المجذاف وهو سهيل | من السفينة | ج/ا/لو — ٣/١/٣٦ | عه/٠ — ٧٥/٠ | جنوب | ا | لي | قاطع |
| صج — ٩٣ | النيّر من الاثنين المضيئين وهو الفرد وعنق الشجاع | من الشجاع | د/يد/كو — ٤/١٤/٢٦ | ك/ل — ٢٠/٣٠ | جنوب | ب | ره | — |
| صد — ٩٤ | جناح الغراب المتقدّم الايمن | — | ه/كز/مو — ٥/٢٧/٤٦ | يد/ن — ١٤/٥٠ | جنوب | ج | لخ | — |
| صه — ٩٥ | الذي على طرف الرجل وهو مشترك له وللشجاع | — | و/د/نح — ٦/٤/٥٨ | يو/ي — ١٦/١٠ | جنوب | ج | لخ | — |
| صو — ٩٦ | الذي على طرف الرجل اليمنى قدّام وهو الوزن | من قنطورس | ز/كب/مو — ٧/٢٢/٤٦ | ما/ي — ٤١/١٠ | جنوب | ا | هي | — |
| صز — ٩٧ | الذي على رجل الركبة اليسرى وهو حضار | — | ز/ح/لو — ٧/٨/٣٦ | مه/ك — ٤٥/٢٠ | جنوب | بك | هي | — |
| صح — ٩٨ | الذي في وسط رأس المجمرة | المجمرة | ح/ي/لو — ٨/١٠/٣٦ | كو/ل — ٢٦/٣٠ | جنوب | دك | ل | — |
| صط — ٩٩ | المقدّم من خارج من القوس الجنوبية ⟩مضعّف؟⟨ | من الاكليل الجنوبي | ح/كج/لو — ٨/٢٣/٣٦ | يا/ل — ١١/٣٠ | جنوب | د | لد | — |
| ق — ١٠٠ | المقدّم من الثلثة وهو الذي على طرف الذنب | من الحوت الجنوبي | ي/ي/كو — ١٠/١٠/٢٦ | كب/يه — ٢٢/١٥ | جنوب | جص | ري | — |

## 10.4 English translation made from the Arabic, with collation against Rosenfeld

**Heading:** “Positions of the fixed stars at the beginning of the first intercalary year of the Malikī [reckoning], which is the year ⟨…⟩ of the Rūmī [era] ⟨…⟩ and 448 (تمح) of Yazdgerd” (= 1079/80 CE).

In the temperament column the Arabic abbreviation of the manuscript is given first, followed by its expansion. The final column records the divergences from Rosenfeld’s Russian translation; an em dash indicates that the two witnesses agree.

| No. | Folio | English translation of the Arabic entry | Coordinates | Mag. | Temperament (MS) | Effect | Proposed modern identification | Note on the collation with Rosenfeld |
|---|---|---|---|---|---|---|---|---|
| 1 | 224r | The brighter of the two Farqadān (the Two Calves) | S4 (Leo); λ 1°56′; β 72°50′ N | 2 | له = Saturn + Venus | Favourable | β UMi (Kochab) | Rosenfeld’s лх = له, i.e. Saturn + Venus. |
| 2 | 224r | The other of the two Farqadān | S4 (Leo); λ 10°36′; β 74°50′ N | 3 | هد = Venus + Mercury | Favourable | γ UMi (Pherkad) | — |
| 3 | 224r | The one at the tip of the tail; it is al-Jady (the Kid) | S2 (Gemini); λ 14°36′; β 66°0′ N | 3 | ل = Saturn | Favourable | α UMi (Polaris) | — |
| 4 | 224r | The one on the back, of those in the quadrilateral — after the correction (baʿd al-iṣlāḥ) | S4 (Leo); λ 14°6′; β 49°0′ N | 2 | خ = Mars | Favourable | α UMa (Dubhe) | MS marginal gloss بعد الاصلاح ('after the correction'), not reflected in Rosenfeld's Russian. Longitude 134°06′ remains ~12° larger than the computed position of α UMa. Rosenfeld's x = خ (Mars), not ه (Venus). |
| 5 | 224r | The one on its flank (al-marāqq) | S4 (Leo); λ 6°36′; β 45°30′ N | 3 (greater) | خ = Mars | Favourable | β UMa (Merak) | MS latitude 30°45= مه/ل′; the 1961 Russian edition prints 45°0′. |
| 6 | 224r | The one at the root of its tail | S4 (Leo); λ 17°36′; β 51°0′ N | 3 (smaller) | خ = Mars | Favourable | δ UMa (Megrez) | — |
| 7 | 224r | The remaining one of them, on the left hind thigh | S4 (Leo); λ 17°26′; β 46°30′ N | 3 (smaller) | خ = Mars | Favourable | γ UMa (Phecda) | — |
| 8 | 224r | The first of the three that follow the root of the tail; it is al-Jawn (the Black Horse) | S4 (Leo); λ 26°36′; β 53°30′ N | 2 | هد = Venus + Mercury | Favourable | ε UMa (Alioth, al-Jawn) | Decisive divergence: MS reads د/كو/لو = Leo 26°36′ (λ 146°36′); Rosenfeld prints 6°36′. The computed position of ε UMa for 1080 CE is 146°05′, confirming the manuscript reading; the printed value derives from misreading )26( كو as )6( و. |
| 9 | 224r | The middle one of them; it is al-ʿAnāq (the She-Goat) | S5 (Virgo); λ 2°26′; β 55°40′ N | 2 | هد = Venus + Mercury | Favourable | ζ UMa (Mizar, al-ʿAnāq) | — |
| 10 | 224r | The third, the one at the tip of the tail; it is al-Qāʾid (the Leader) | S5 (Virgo); λ 14°16′; β 54°0′ N | 2 | هد = Venus + Mercury | Favourable | η UMa (Alkaid, al-Qāʾid) | — |
| 11 | 224r | One of the two stars of ⟨…⟩, the one inclining toward the north | S5 (Virgo); λ 22°46′; β 84°50′ N | 3 | لخ = Saturn + Mars | Favourable | ζ Dra (Aldhibah) | Constellation word partly effaced; Rosenfeld renders simply 'one of the two stars deviating northwards'. |
| 12 | 224r | The northern of the two on the side toward the west, [inclining] to the north | S5 (Virgo); λ 24°6′; β 78°0′ N | 3 | لخ = Saturn + Mars | Favourable | η Dra | — |
| 13 | 224r | The one touching above the right shoulder, one of the two western stars | S0 (Aries); λ 1°6′; β 69°0′ N | 3 | له = Saturn + Venus | Favourable | α Cep (Alderamin) | — |
| 14 | 224r | The one between the thighs of al-ʿAwwāʾ; it is al-Simāk al-Rāmiḥ | S6 (Libra); λ 11°26′; β 31°30′ N | 1 | يخ = Jupiter + Mars | Favourable | α Boo (Arcturus) | — |
| 15 | 224r | The bright one ⟨…⟩; it is the bright star of al-Fakka | S6 (Libra); λ 29°6′; β 44°30′ N | 2 | هد = Venus + Mercury | Favourable | α CrB (Alphecca) | First epithet uncertain in the MS; Rosenfeld: 'the brightest in the Crown'. |
| 16 | 224r | The one on the head; it is Kalb al-Rāʿī (the Shepherd's Dog) | S8 (Sagittarius); λ 2°6′; β 37°30′ N | 3 (smaller) | هد = Venus + Mercury | Favourable | α Her (Rasalgethi) | — |
| 17 | 224r | The one on ⟨the shell⟩; it is al-Nasr al-Wāqiʿ (the Falling Eagle), also called al-Lūrā (the Lyre) | S9 (Capricorn); λ 1°46′; β 62°0′ N | 1 | هد = Venus + Mercury | Favourable | α Lyr (Vega) | The word rendered 'shell' is unclear in the MS; the Arabic explicitly adds the Greek loan اللورا (al-Lūrā, 'the Lyre'), which Rosenfeld also preserves. |
| 18 | 224r | The one on the mouth; it is Minqār al-Dajāja (the Hen's Beak) | S9 (Capricorn); λ 18°56′; β 49°20′ N | 3 (smaller) | هد = Venus + Mercury | Favourable | β Cyg (Albireo) | — |
| 19 | 224r | The bright one in the tail; it is al-Ridf (the Rear One) | S10 (Aquarius); λ 23°36′; β 60°0′ N | 2 | هد = Venus + Mercury | Favourable | α Cyg (Deneb) | — |
| 20 | 224r | The one in the middle of the throne, on the camel's hump; it is al-Kaff al-Khaḍīb (the Dyed Hand) | S0 (Aries); λ 22°16′; β 51°40′ N | 3 | له = Saturn + Venus | Favourable | β Cas (Caph) | — |
| 21 | 224r | The northern nebulous one at the end of the right hand; it is Miʿṣam al-Thurayyā (the Wrist of the Pleiades) | S1 (Taurus); λ 11°6′; β 40°30′ N | nebulous | خد = Mars + Mercury | Favourable | h + χ Per (NGC 869/884) | The Arabic reads ‘northern nebulous’, matching Rosenfeld; the MS magnitude cell has سحابي (‘nebulous’) rather than a numeral. |
| 22 | 224r | The bright one on the right side; it is Mirfaq al-Thurayyā (the Elbow of the Pleiades) | S1 (Taurus); λ 19°16′; β 30°0′ N | 2 | خد = Mars + Mercury | Favourable | α Per (Mirfak) | — |
| 23 | 224r | The bright one in Raʾs al-Ghūl (the Head of the Gorgon) | S1 (Taurus); λ 14°6′; β 23°0′ N | 2 (smaller) | لد = Saturn + Mercury | Favourable | β Per (Algol) | — |
| 24 | 224r | The one on the left shoulder; it is al-ʿAyyūq | S2 (Gemini); λ 9°26′; β 22°30′ N | 1 | خد = Mars + Mercury | Favourable | α Aur (Capella) | — |
| 25 | 224r | The one on the right shoulder; it is one of the aʿlām (marks) and what follows al-Najm (the Pleiades) | S2 (Gemini); λ 17°16′; β 20°0′ N | 2 | خد = Mars + Mercury | Favourable | β Aur (Menkalinan) | Arabic اعلام (‘marks, signs’); Rosenfeld renders ‘one of the signs of the Pleiades’. |
| 26 | 224r | The one on the right ankle, shared between it and the northern horn of Taurus | S2 (Gemini); λ 10°6′; β 5°0′ N | 2 | خد = Mars + Mercury | Favourable | β Tau (Elnath) | — |
| 27 | 224r | The one on the head; it is al-Rāʿī (the Shepherd) | S8 (Sagittarius); λ 9°16′; β 36°0′ N | 3 | له = Saturn + Venus | Favourable | α Oph (Rasalhague) | — |
| 28 | 224r | The preceding of the two on the right shoulder; it is Kulayb al-Rāʿī (the Shepherd’s Little Dog) | S8 (Sagittarius); λ 12°26′; β 27°15′ N | 3 (smaller) | له = Saturn + Venus | Favourable | β Oph (Cebalrai) | — |
| 29 | 224r | The middle of the three on the serpent’s neck in the Syrian (northern) row; it is ʿUnuq al-Ḥayya | S7 (Scorpio); λ 8°46′; β 25°20′ N | 3 | له = Saturn + Venus | Favourable | α Ser (Unukalhai) | — |
| 30 | 224r | The solitary one on the arrowhead | S9 (Capricorn); λ 24°36′; β 39°20′ N | 3 | خه = Mars + Venus | Favourable | γ Sge | — |
| 31 | 224r | The bright one between the shoulders; it is al-Nasr al-Ṭāʾir (the Flying Eagle) | S9 (Capricorn); λ 18°16′; β 29°10′ N | 2 (greater) | خي = Mars + Jupiter | Favourable | α Aql (Altair) | — |
| 32 | 224r | The preceding of the three in the tail; it is Dhanab al-Dulfīn (the Dolphin’s Tail) | S10 (Aquarius); λ 2°6′; β 29°10′ N | 4 (greater) | خه = Mars + Venus | Favourable | ε Del | — |
| 33 | 224r | The preceding of the two in the head | S10 (Aquarius); λ 10°46′; β 20°30′ N | 4 | ل = Saturn | Favourable | α Equ (Kitalpha) | — |
| 34 | 224r | The one on the navel, shared between it and the head of al-Marʾa al-Musalsala (Andromeda) | S0 (Aries); λ 2°16′; β 26°0′ N | 2 (smaller) | خد = Mars + Mercury | Favourable | α And (Alpheratz) | — |
| 35 | 224v | The one on the back and at the tip of the wing; it is Janāḥ al-Faras (the Horse’s Wing) | S11 (Pisces); λ 26°36′; β 12°30′ N | 2 (smaller) | خد = Mars + Mercury | Unfavourable (قاطع) | γ Peg (Algenib) | — |
| 36 | 224v | The one on the right shoulder and the beginning of the leg from it; it is Mankib al-Faras | S11 (Pisces); λ 16°36′; β 31°0′ N | 2 (smaller) | خد = Mars + Mercury | Unfavourable (قاطع) | β Peg (Scheat) | — |

| 37 | 224v | The one between the shoulder blades and the shoulder of the wing; it is Matn al-Faras | S11 (Pisces); λ 11°6′; β 19°40′ N | 2 (smaller) | خد = Mars + Mercury | Favourable | α Peg (Markab) | — |
|---|---|---|---|---|---|---|---|---|
| 38 | 224v | The one on the muzzle; it is Jaḥfalat al-Faras | S10 (Aquarius); λ 19°46′; β 22°30′ N | 3 (smaller) | خد = Mars + Mercury | Favourable | ε Peg (Enif) | — |
| 39 | 224v | The southern of the three above the back; it is Baṭn al-Ḥūt and al-Rishāʾ | S0 (Aries); λ 18°16′; β 26°20′ N | 2 (smaller) | ه = Venus | Favourable | β And (Mirach) | Rosenfeld's x = ه (Venus). |
| 40 | 224v | The one at the apex of the Triangle; it is one of ⟨…⟩ | S0 (Aries); λ 25°26′; β 16°30′ N | 3 | د = Mercury | Favourable | α Tri (Mothallah) | Final word of the Arabic uncertain; Rosenfeld: 'one of the Two Friends'. |
| 41 | 224v | The preceding of the two on the horn | S0 (Aries); λ 21°6′; β 7°20′ N | 3 (smaller) | خل = Mars + Saturn | Favourable | γ Ari (Mesarthim) | — |
| 42 | 224v | The following one of them | S0 (Aries); λ 22°6′; β 8°20′ S | 3 | خل = Mars + Saturn | Favourable | β Ari (Sheratan) lat. N | MS الجهات column has جنوب ('south'), as in Rosenfeld, although β Ari in fact has northern latitude; with 'north' the coordinates agree with β Ari to ~1°. |
| 43 | 224v | The one above the head; it is al-Nāṭiḥ (the Butting One) | S0 (Aries); λ 25°6′; β 10°0′ S | 3 (greater) | خل = Mars + Saturn | Unfavourable (قاطع) | α Ari (Hamal) lat. N | As for no. 42: the MS gives جنوب, but the position corresponds to α Ari with northern latitude. |
| 44 | 224v | The bright one tending to redness, figured in [the shape of] the letter dāl, on the southern eye; it is al-Dabarān | S1 (Taurus); λ 27°6′; β 5°10′ S | 1 | خ = Mars | Unfavourable (قاطع) | α Tau (Aldebaran) | The Arabic specifies المصوّر في الدال ('figured in the letter dāl'), i.e. the Hyades V-shape; Rosenfeld renders 'in the figure of the letter dāl'. Rosenfeld's x = خ (Mars). |
| 45 | 224v | The one on the head of the preceding Twin | S3 (Cancer); λ 6°46′; β 9°40′ N | 2 | )؟(يد = Jupiter + Mercury | Favourable | α Gem (Castor) | Temperament: the MS abbreviation may be read either يد (Jupiter + Mercury, as in Rosenfeld) or خد (Mars + Mercury); the reading is left open here. |
| 46 | 224v | The one tending to redness on the head of the second Twin | S3 (Cancer); λ 11°6′; β 6°15′ N | 2 | خ = Mars | Favourable | β Gem (Pollux) | Rosenfeld's x = خ (Mars). |
| 47 | 224v | The middle nebulous one in the breast; it is al-Miʿlaf (the Manger) and al-Nathra | S3 (Cancer); λ 24°46′; β 0°40′ N | nebulous | خس = Mars + Sun | Unfavourable (قاطع) | M 44 (Praesepe) | MS خس = Mars + Sun. |
| 48 | 224v | The northern of the two in the head; it is Hāmat al-Asad (the Lion's Crown) | S4 (Leo); λ 8°46′; β 12°0′ N | 3 (smaller) | لخ = Saturn + Mars | Favourable | μ Leo (Rasalas) | — |
| 49 | 224v | The fourth, the middle of the three; it is ⟨…⟩ of the Lion | S4 (Leo); λ 8°36′; β 8°30′ N | 1 | يخ = Jupiter + Mars | Favourable | ε Leo (?) | A word of the Arabic epithet is unclear; Rosenfeld: 'the Shoulder of the Beneficent Lion'. The magnitude (1) and the coordinates do not agree with one another; see the apparatus. |
| 50 | 224v | The one on the heart, called al-Malikī (the Royal) and al-Falakī; it is Qalb al-Asad | S4 (Leo); λ 16°56′; β 0°10′ N | 1 | يخ = Jupiter + Mars | Favourable | α Leo (Regulus) | — |
| 51 | 224v | The following one of them; it is one of the Lion's ⟨knees⟩ | S4 (Leo); λ 28°36′; β 13°40′ N | 2 | له = Saturn + Venus | Favourable | δ Leo (Zosma) | Arabic uncertain; Rosenfeld: 'the rear one of the two on the loin'. |
| 52 | 224v | The lower of the two at the end of ⟨the hindquarters⟩; it belongs to al-Zubra | S5 (Virgo); λ 0°46′; β 9°40′ N | 3 | له = Saturn + Venus | Favourable | θ Leo (Chort) | Arabic partly unclear; Rosenfeld: 'the lower one of the two on the buttocks, deviating southward'. |
| 53 | 224v | The one at the tip of the tail; it is Dhanab al-Asad, Ṣulb al-Asad and al-Ṣarfa | S5 (Virgo); λ 8°56′; β 11°50′ N | 1 | له = Saturn + Venus | Favourable | β Leo (Denebola) | — |
| 54 | 224v | The one at the tip of the southern left wing; it is the first of the stars of al-ʿAwwāʾ | S5 (Virgo); λ 13°26′; β 0°10′ N | 3 | دخ = Mercury + Mars | Favourable | β Vir (Zavijava) | — |
| 55 | 224v | The one on the left palm, called al-Sunbula (the Ear of | S6 (Libra); λ 11°6′; β 2°0′ S | 1 (smaller) | ده = Mercury + Mars | Favourable | α Vir (Spica) | Temperament in the MS reads ده (Mercury + Venus); Rosenfeld's дх leaves the second letter ambiguous. |
| | | Grain); it is al-Simāk al-Aʿzal | | | | | | |
| 56 | 224v | The brighter of the two at the tip of the southern claw (al-zubānā) | S7 (Scorpio); λ 2°26′; β 0°40′ N | 3 (greater) | يد = Jupiter + Mercury | Favourable | α² Lib (Zubenelgenubi) | — |
| 57 | 224v | The brighter of the two at the tip of the northern claw (al-zubānā) | S7 (Scorpio); λ 6°36′; β 8°50′ N | 3 (greater) | يد = Jupiter + Mercury | Unfavourable (قاطع) | β Lib (Zubeneschamali) | — |
| 58 | 224v | The middle of the three, the one tending to redness; it is Qalb al-ʿAqrab | S7 (Scorpio); λ 27°6′; β 4°0′ S | 2 | خي = Mars + Jupiter | Unfavourable (قاطع) | α Sco (Antares) | — |
| 59 | 224v | The following of the two in the sting, of al-Shawla | S8 (Sagittarius); λ 11°56′; β 13°20′ S | 3 | (لخ؟) = Saturn + Mars | Favourable | λ Sco (Shaula) | Temperament reading uncertain in the MS (لخ or له). |
| 60 | 224v | The preceding of the two in the sting, of al-Shawla | S8 (Sagittarius); λ 11°26′; β 13°30′ S | 3 (smaller) | (له؟) = Saturn + Mars | Favourable | υ Sco (Lesath) | Temperament reading uncertain in the MS (له or لخ). |
| 61 | 224v | The nebulous one following the sting; it is ⟨…⟩ of al-Shawla | S8 (Sagittarius); λ 15°36′; β 13°30′ S | 4 (smaller) | خد = Mars + Mercury | Unfavourable (قاطع) | M 7 (Ptolemy Cluster) | Final epithet unclear; Rosenfeld: ‘the nebulous one following the sting’. The object is the cluster M7. |
| 62 | 224v | The one on the blade of the arrow | S8 (Sagittarius); λ 18°56′; β 6°30′ S | 3 (smaller) | لر = Saturn + Moon | Unfavourable (قاطع) | γ² Sgr (Alnasl) | — |
| 63 | 224v | The one in the grip of the left hand | S8 (Sagittarius); λ 22°6′; β 10°30′ S | 3 | يد = Jupiter + Mercury | Favourable | ε Sgr (Kaus Australis) | — |
| 64 | 224v | The one on the southern side of the bow | S8 (Sagittarius); λ 22°26′; β 10°50′ S | 3 (greater) | يد = Jupiter + Mercury | Favourable | δ/ε Sgr (?) | — |
| 65 | 224v | The one on the right foreleg in front, the southern of the two southern ones | S8 (Sagittarius); λ 21°6′; β 13°0′ S | 3 (smaller) | يد = Jupiter + Mercury | Unfavourable (قاطع) | η Sgr | One word of the Arabic is unclear; Rosenfeld: ‘the one at the right front hoof’. |
| 66 | 224v | The doubled nebulous one on the eye; it is ʿAyn al-Rāmī (the Archer’s Eye) | S8 (Sagittarius); λ 29°36′; β 0°45′ N | nebulous | در = Mercury + Moon | Unfavourable (قاطع) | ν¹/ν² Sgr (ʿAyn al-Rāmī) | Arabic المصعّف (‘doubled’), i.e. a double star; Rosenfeld renders ‘the northern double one in the eye’. |
| 67 | 224v | The one on the left shoulder, the following of the two northern ones | S8 (Sagittarius); λ 29°46′; β 3°10′ S | 3 (smaller) | يد = Jupiter + Mercury | Favourable | σ Sgr (Nunki) | — |
| 68 | 224v | The one preceding them both, shared in ⟨the shaft of the arrow⟩ | S8 (Sagittarius); λ 27°26′; β 3°50′ S | 4 (greater) | يد = Jupiter + Mercury | Favourable | φ Sgr | Final word read tentatively as القدح (‘the shaft of the arrow’); Rosenfeld: ‘the upper part of the arrow’. |
| 69 | 225r | The middle one of them, on the shoulder | S9 (Capricorn); λ 2°6′; β 4°30′ S | 4 (greater) | يد = Jupiter + Mercury | Favourable | τ Sgr | — |
| 70 | 225r | ⟨The third⟩, the one beneath the armpit | S9 (Capricorn); λ 0°46′; β 6°45′ S | 3 | يد = Jupiter + Mercury | Favourable | ζ Sgr (Ascella) | — |
| 71 | 225r | The northern of the three on the following horn | S9 (Capricorn); λ 21°46′; β 7°20′ N | 3 (smaller) | هخ = Venus + Mars | Favourable | α Cap (Algedi) | MS هخ = Venus + Mars. |
| 72 | 225r | The southern of the three | S9 (Capricorn); λ 21°46′; β 5°0′ N | 3 (smaller) | هخ = Venus + Mars | Favourable | β Cap (Dabih) | — |

| 73 | 225r | The one on the left shoulder; it belongs to Saʿd al-Suʿūd | S10 (Aquarius); λ 10°56′; β 8°50′ N | 3 (smaller) | لد = Saturn + Mercury | Favourable | β Aqr (Sadalsuud) | — |
|---|---|---|---|---|---|---|---|---|
| 74 | 225r | The one at the end of the water, on the mouth of the Southern Fish; it is al-Ḍifdaʿ al-Awwal (the First Frog) | S10 (Aquarius); λ 21°26′; β 23°0′ S | 1 | (ره؟) = Moon + Venus | Favourable | α PsA (Fomalhaut) | Temperament reading uncertain (ره). |
| 75 | 225r | The one in the mouth of the preceding Fish | S11 (Pisces); λ 6°6′; β 9°15′ N | 4 | دل = Mercury + Saturn | Favourable | β Psc (Fum al-Samaka) | — |
| 76 | 225r | The one at the knot of the two cords | S0 (Aries); λ 16°56′; β 8°30′ S | 3 (smaller) | دل = Mercury + Saturn | Favourable | α Psc (Alrescha) | — |
| 77 | 225r | The one on the northern branch of the two in ⟨the fish of⟩ the tail; it is Dhanab Qayṭus | S11 (Pisces); λ 18°46′; β 9°40′ S | 3 (smaller) | ل = Saturn | Favourable | ι Cet (Deneb Kaitos Shemali) | — |
| 78 | 225r | The one on the southern branch of it, of the tail; it is al-Ḍifdaʿ al-Thānī (the Second Frog) | S11 (Pisces); λ 20°6′; β 20°20′ S | 3 (greater) | ل = Saturn | Favourable | β Cet (Diphda) | — |
| 79 | 225r | ⟨The nebulous one⟩ in the head of al-Jabbār; it is al-Haqʿa and Raʾs al-Jabbār | S2 (Gemini); λ 11°26′; β 13°50′ S | nebulous | خد = Mars + Mercury | Unfavourable (قاطع) | λ Ori (Meissa) | First word partly effaced; the magnitude column gives سحابي, and Rosenfeld likewise marks the entry as nebulous. |
| 80 | 225r | The bright one on the right shoulder; it is Yad al-Jawzāʾ and Mankib al-Jawzāʾ | S2 (Gemini); λ 16°26′; β 17°0′ S | 1 (smaller) | خد = Mars + Mercury | Unfavourable (قاطع) | α Ori (Betelgeuse) | — |
| 81 | 225r | The one on the left shoulder; it is al-Nājidh and al-Mirzam | S2 (Gemini); λ 8°26′; β 17°30′ S | 2 | يل = Jupiter + Saturn | Favourable | γ Ori (Bellatrix) | — |
| 82 | 225r | The preceding of the three on the belt | S2 (Gemini); λ 9°46′; β 24°10′ S | 2 | يل = Jupiter + Saturn | Favourable | δ Ori (Mintaka) | — |
| 83 | 225r | The middle one of them | S2 (Gemini); λ 11°46′; β 24°50′ S | 2 | يل = Jupiter + Saturn | Favourable | ε Ori (Alnilam) | — |
| 84 | 225r | The following of the three | S2 (Gemini); λ 12°36′; β 25°40′ S | 2 | يل = Jupiter + Saturn | Favourable | ζ Ori (Alnitak) | — |
| 85 | 225r | The bright one on the left foot; it is Rijl al-Jawzāʾ al-Yusrā and Rāʿī al-Jawzāʾ | S2 (Gemini); λ 4°16′; β 31°30′ S | 1 | يل = Jupiter + Saturn | Unfavourable (قاطع) | β Ori (Rigel) | — |
| 86 | 225r | The bright one at the end of the River; it is al-Ẓalīm (the Ostrich) | S0 (Aries); λ 14°36′; β 53°30′ S | 1 | يه = Jupiter + Venus (?) | Favourable | θ Eri (Acamar) | MS يه = Jupiter + Venus. |
| 87 | 225r | The one in the middle of the body | S2 (Gemini); λ 10°16′; β 41°30′ S | 3 (smaller) | لخ = Saturn + Mars | Favourable | α Lep (Arneb) | — |
| 88 | 225r | The one on the mouth; it is al-Shiʿrā al-Yamāniyya, al-ʿAbūr and Kalb al-Jabbār | S3 (Cancer); λ 2°6′; β 39°10′ S | 1 | يخ = Jupiter + Mars | Favourable | α CMa (Sirius) | — |
| 89 | 225r | The one at the end of the preceding right foot; it is al-Mirzam | S2 (Gemini); λ 25°26′; β 41°20′ S | 3 | يخ = Jupiter + Mars | Favourable | β CMa (Mirzam) | — |
| 90 | 225r | The one on the neck; it is al-Mirzam | S3 (Cancer); λ 9°26′; β 14°0′ S | 4 | خد = Mars + Mercury | Favourable | β CMi (Gomeisa) | The MS repeats المرزم for both nos. 89 and 90; Rosenfeld renders no. 90 as ‘the Tether’. |

| 91 | 225r | The bright one behind ⟨…⟩; it is al-Shiʿrā al-Shāmiyya and al-Ghumayṣāʾ | S3 (Cancer); λ 13°36′; β 16°10′ S | 1 | خد = Mars + Mercury | Favourable | α CMi (Procyon) | One word of the Arabic is unclear; Rosenfeld: ‘the bright one behind, i.e. the Syrian Sirius, the Weeping One’. |
|---|---|---|---|---|---|---|---|---|
| 92 | 225r | The preceding of the two on the oar; it is Suhayl | S3 (Cancer); λ 1°36′; β 75°0′ S | 1 | لي = Saturn + Jupiter | Unfavourable (قاطع) | α Car (Canopus, Suhayl) | — |
| 93 | 225r | The brighter of the two shining ones; it is al-Fard and ʿUnuq al-Shujāʿ | S4 (Leo); λ 14°26′; β 20°30′ S | 2 | زه = Moon + Venus | Favourable | α Hya (Alphard) | — |
| 94 | 225r | The preceding right wing of the Crow | S5 (Virgo); λ 27°46′; β 14°50′ S | 3 | لخ = Saturn + Mars | Favourable | γ Crv (Gienah) | — |
| 95 | 225r | The one at the end of the foot, shared between it and Hydra | S6 (Libra); λ 4°58′; β 16°10′ S | 3 | لخ = Saturn + Mars | Favourable | β Crv (Kraz) | — |
| 96 | 225r | The one at the end of the right foreleg in front; it is al-Wazn | S7 (Scorpio); λ 22°46′; β 41°10′ S | 1 | هي = Venus + Jupiter | Favourable | α Cen (?) | Arabic الوزن; Rosenfeld transliterates ‘Wazn’. The identification with α Cen is proposed on the coordinates and the magnitude. |
| 97 | 225r | The one on the left knee; it is Ḥaḍār | S7 (Scorpio); λ 8°36′; β 45°20′ S | 2 (greater) | هي = Venus + Jupiter | Favourable | β Cen (Hadar) | — |
| 98 | 225r | The one in the middle of the top of the Censer | S8 (Sagittarius); λ 10°36′; β 26°30′ S | 4 (greater) | ل = Saturn | Favourable | α Ara | — |
| 99 | 225r | The foremost external one on the southern arc ⟨…⟩ | S8 (Sagittarius); λ 23°36′; β 11°30′ S | 4 | لد = Saturn + Mercury | Favourable | α Tel (α Telescopii) | The description corresponds to Ptolemy’s first star of Corona Australis, the ‘foremost [or preceding] external one on the southern arc’, which is identified with α Telescopii. The star belonged to Corona Australis in the Ptolemaic constellation scheme but lies in the modern constellation Telescopium. The transmitted latitude, β = −11°30′, is anomalous; the corresponding Ptolemaic latitude is approximately −21°30′, suggesting corruption of the latitude in the transmitted table. |
| 100 | 225r | The foremost of the three, the one at the tip of the tail | S10 (Aquarius); λ 10°26′; β 22°15′ S | 3 (smaller) | زي = Moon + Jupiter | Favourable | μ PsA (?) | Rosenfeld–Yushkevich identify the star with ι Piscis Austrini; the coordinates agree better with μ PsA. |

### *10.4.1 The principal divergences from the Russian translation*

The collation yields one substantive correction of the printed edition, several readings that Rosenfeld’s Cyrillic transcription could not express, and a small number of places where the Arabic is more explicit than the Russian.

1. **Entry 8 (al-Jawn, ε UMa).** The manuscript reads د/كو/لو, that is Leo 26°36′ (λ 146°36′); the 1961 edition prints 6°36′. Precessing the modern position of ε Ursae Majoris back to 1080 CE gives λ 146°05′, so the manuscript reading is correct to about half a degree, whereas the printed value is some 20° in error. The corruption is easily explained: )26( كو was read as )6( و.
2. **Entry 5.** The manuscript gives the latitude as 30°45= مه/ل′, where the 1961 edition prints 45°0′.
3. **The temperament column.** Rosenfeld’s Cyrillic x conflates خ (Mars) and ه (Venus), so that his transcription cannot distinguish, for example, entry 4 (خ, Mars) from entry 39 (ه, Venus), or entry 1 (له, Saturn + Venus) from entry 11 (لخ, Saturn + Mars). All 100 abbreviations are restored here from the manuscript; four readings (entries 45, 59, 60 and 74) remain uncertain and are marked as such.
4. **Entry 4.** The manuscript carries the marginal note بعد الاصلاح (“after the correction”), which the Russian translation does not reproduce. The entry is also the one whose transmitted longitude (134°06′) departs most widely from the computed position of α UMa (122°21′), so that the gloss and the anomaly may well be connected.
5. **Entries 42 and 43.** Both are marked جنوب (“south”) in the direction column of the manuscript, as in the Russian translation, although β and α Arietis have northern ecliptic latitude; read as northern, the transmitted coordinates agree with those stars to within about one degree.
6. **Wording.** In a number of entries the Arabic is more specific than the Russian rendering: entry 44 describes the star as المصوّر في الدال, “figured in [the shape of] the letter dāl”; entry 66 calls the object المصعّف, “doubled”; entry 17 adds the Greek loan-name اللورا beside النسر الواقع. These are recorded in the table.

Apart from these points, the Arabic text and the Russian translation agree closely: the numerical columns of all three folios were read in full and, with the two exceptions noted above, coincide with the values printed in 1961. The collation therefore confirms the general reliability of Rosenfeld’s edition while showing why an edition made directly from the manuscript is still required.

## 10.5 Commentary derived from Rosenfeld and Yushkevich

Rosenfeld and Yushkevich identify *Dustūr al-munajjimīn* as a compilation of ten books. In their description the stellar astronomy section occupies fols. 215a–231a, while the Malikshāhī catalogue occupies fols. 224r–225r (224a–225a in the a/b foliation). They also note that tables of 36 stars for the years 1080, 1110, 1140 and 1170 follow the 100-star catalogue [6]. Their older assumption that the surviving physical manuscript itself was a twelfth-century product must now be corrected in the light of the later palaeographical and provenance research discussed above.

Their principal astronomical comments may be rendered as follows:

| R&Y note | English rendering |
|---|---|
| 2 | The Malikī intercalary reckoning refers to the Malikī calendrical system; Rosenfeld and Yushkevich date its beginning to March 1079. |
| 3 | The “1490 Rūmī year” is interpreted as 1078/79 and as an instance of the Seleucid or Alexander era. |
| 4 | Year 448 of Yazdgerd corresponds to 1079/80. |
| 5 | Stellar longitudes are expressed by zodiacal signs, degrees and minutes; for 96 of 100 stars the longitude exceeds Ptolemy’s by 14°26′. |
| 6 | Stellar latitudes are given in degrees and minutes; 87 of the 100 values agree with Ptolemy. |
| 7 | Magnitudes follow the Ptolemaic scale. The manuscript’s *kāf* and *ṣād* indicate greater or smaller relative magnitude; Rosenfeld renders these by **б** and **м**. |
| 8 | The column of *mizājāt* (“temperaments”) probably reflects medieval astrological ideas about stellar influences. **Rosenfeld and Yushkevich state that the one- or two-letter *mizāj* abbreviations were not fully understood by them. The present study goes beyond their interpretation by collating these signs directly with the manuscript and identifying their planetary values.** |
| 9 | The effects classified as *salīm* and *qāṭiʿ* are rendered respectively as favourable and unfavourable. |
| 10 | Ursa Minor is Arabic *al-dubb al-aṣghar*, “the lesser bear”. The three entries are identified with β, γ and α Ursae Minoris, the last being Polaris. |
| 11 | Ursa Major is *al-dubb al-akbar*. Rosenfeld and Yushkevich relate the listed stars to the familiar principal stars of the Great Bear and discuss Arabic names connected with the bear, its body and the traditional funeral-procession imagery. |
| 12 | Draco is *al-tinnīn*; the two entries are identified with ζ and η Draconis. |
| 13 | Cepheus appears in a form derived from the Greek name; the listed star is identified with α Cephei (Alderamin). |
| 14 | Boötes is *al-ʿawwāʾ*; the listed star is Arcturus. The expression Simāk is discussed in connection with the Arabic stellar-name tradition. |
| 15 | Corona Borealis is *al-iklīl*, “the crown”, also known as *al-fakka*, the “Beggars’ Bowl”; the star is α Coronae Borealis, Alphecca. |
| 16 | Hercules is described through the traditional “kneeling” figure; the star is α Herculis, Ras Algethi. |
| 17 | Lyra is associated with the tortoise and with the Falling Eagle; the star is α Lyrae, Vega. |
| 18 | Cygnus is *al-dajāja*, “the hen”; the entries are associated with β Cygni and α Cygni, Deneb. |
| 19 | Cassiopeia is “the possessor of the throne”; the listed star is β Cassiopeiae, Caph. |
| 20 | The Perseus entries are identified with the traditional stellar complex including Mirfak and Algol; the latter name is connected with Arabic *al-ghūl*. |
| 21 | Auriga is described as the “holder of the reins”; the entries include Capella and neighbouring bright stars. |
| 22 | Ophiuchus is *al-ḥawwāʾ*, the serpent-charmer; the stars include Ras Alhague and Cebalrai. |
| 23 | Serpens is *al-ḥayya*. Rosenfeld and Yushkevich explain the “Yemeni” row as the southern/front part of the traditional Serpent figure. |
| 24 | Sagitta is *al-sahm*, “the arrow”; the listed star is identified with γ Sagittae. |
| 25 | Aquila is *al-ʿuqāb*; the listed star is Altair, from the Arabic expression meaning “the flying eagle”. |
| 26 | Delphinus preserves a name transliterated from Greek; the entry is identified with ε Delphini. |
| 27 | Equuleus is represented by an Arabic expression meaning a part of the horse; the listed star is α Equulei. |
| 28 | Pegasus is “the greater horse”; the entries are identified with α Andromedae and several principal stars of Pegasus. |
| 29 | Andromeda is *al-musalsala*, “the chained one”; the listed star is β Andromedae. |

| 30 | Triangulum is *al-muthallath*, “the triangle”; the star is α Trianguli. |
|---|---|
| 31 | Aries is *al-ḥamal*, “the ram/lamb”; the three stars include γ, β and α Arietis (Hamal). |
| 32 | Taurus is *al-thawr*, “the bull”; the entry is Aldebaran, whose Arabic name refers to following the Pleiades. |
| 33 | Gemini is *al-tawʾamān*, “the twins”; the two stars are Castor and Pollux. |
| 34 | Cancer is *al-saraṭān*; the listed object is identified with ε Cancri. **Editorial note: Rosenfeld–Yushkevich’s ε Cnc identification is retained here as part of their commentary; in Table 2 the present study instead identifies the explicitly nebulous entry with Praesepe (M44/NGC 2632).** |
| 35 | Leo is *al-asad*. The entries include Regulus and Denebola; Rosenfeld and Yushkevich discuss the Arabic origin of the latter and the royal terminology associated with Regulus. |
| 36 | Virgo is *al-ʿadhrāʾ*; the entries include Spica, whose Latin name “ear of grain” corresponds to the traditional Arabic imagery. |
| 37 | Libra is *al-mīzān*. The “claws” terminology preserves the older association of Libra’s stars with the claws of Scorpius. |
| 38 | Scorpius is *al-ʿaqrab*. The principal listed star is Antares, distinguished by its reddish appearance. |
| 39 | Sagittarius is *al-rāmī*, “the archer”; several principal stars are identified, including a double-star entry. |
| 40 | Capricornus is *al-jady*, “the kid”; the listed stars correspond to α and β Capricorni. |
| 41 | Aquarius includes a star of Aquarius and Fomalhaut in Piscis Austrinus; the latter name derives from an Arabic expression meaning “mouth of the fish”. |
| 42 | Pisces is “the two fishes”; the entries correspond to β and α Piscium. |
| 43 | Cetus preserves a name ultimately transliterated from Greek. The expression “from Cetus” is reportedly absent in the manuscript at this point; Deneb Kaitos is connected with Arabic “tail of Cetus”. |
| 44 | Orion is *al-jabbār*, “the giant”, and also associated with the difficult traditional term *al-jawzāʾ*. The manuscript reportedly lacks the explicit heading “from Orion”. The listed stars include Betelgeuse, Bellatrix, the belt stars and Rigel. |
| 45 | Eridanus is ‘the river’. Rosenfeld and Yushkevich identify the listed star with θ Eridani and discuss the traditional name Achernar, derived from Arabic *ākhir al-nahr*, ‘the end of the river’. |
| 46 | Lepus is *al-arnab*; the listed star is α Leporis, Arneb. |
| 47 | Canis Major is “the greater dog”. The entries include Sirius and Mirzam; Sirius is characterised as the “Yemeni”, or southern, Sirius. |
| 48 | Canis Minor is “the lesser dog”; the entries include Procyon, historically the “star before the dog”. |
| 49 | Argo Navis is “the ship”; the listed bright star is Canopus, Arabic Suhayl. |
| 50 | Hydra is *al-shujāʿ*; the listed star is Alphard, from Arabic *al-fard*, “the solitary one”. |
| 51 | Corvus is *al-ghurāb*; the entries are identified with two principal stars of Corvus. |
| 52 | Centaurus preserves a Greek-derived name; the entries are α and β Centauri. Rosenfeld and Yushkevich considered the precise meanings of *Wazn* and *Ḥaḍar* uncertain in this context. |
| 53 | Ara is *al-mijmara*; the entry is associated with a principal star of the Altar. |
| 54 | Corona Australis is *al-iklīl al-janūbī*, “the southern crown”; **Rosenfeld and Yushkevich explicitly identify the listed star with α Telescopii, a star historically associated with Corona Australis before the modern constellation boundaries were established.** |
| 55 | Piscis Austrinus is *al-ḥūt al-janūbī*, “the southern fish”; the commentary notes that its α star had already appeared under Aquarius. **The star listed here is identified by Rosenfeld and Yushkevich as ι Piscis Austrini.** |

## 10.6 Critical apparatus

**1. Manuscript extent.** The 100-star catalogue occupies BnF Arabe 5968, fols. 224r–225r: entries 1–34 on fol. 224r, 35–68 on fol. 224v, and 69–100 on fol. 225r.

**2. Temperament abbreviations.** The manuscript column records the planetary “temperament” (mizāj) assigned to each fixed star. The abbreviations use letters associated with the Arabic planet names. A special difficulty is that Rosenfeld’s Cyrillic x can correspond either to خ (Mars) or ه (Venus); the manuscript image is therefore decisive.

**3. Modern identifications.** The modern identifications are editorial aids, not part of the medieval manuscript. Secure identifications are based on the descriptive location within the constellation, the coordinates, traditional Arabic names, and comparison with the Ptolemaic star catalogue. A question mark or cautionary note indicates a genuine uncertainty.

**4. Coordinate updating.** The strong agreement of most latitudes with Ptolemy and the approximately systematic increase of longitudes support the interpretation that much of the catalogue was updated from inherited Ptolemaic data for a later epoch rather than independently re-observed star by star.

**5. Entry 99.** The description corresponds to the first star of Corona Australis in the Ptolemaic catalogue, ‘the foremost [or preceding] external one on the southern arc’, securely identified in modern catalogues with α Telescopii. Although this star belonged to Corona Australis in the Ptolemaic constellation scheme, it was later incorporated into Telescopium. The Malikshāhī longitude is compatible with this identification after epochal updating, whereas the transmitted latitude of −11°30′ is problematic; the corresponding Ptolemaic latitude is approximately −21°30′ and may indicate a scribal or transmission error.

**6. Entry 100.** Rosenfeld–Yushkevich identify the final star with ι Piscis Austrini. The Arabic description, المقدّم من الثلثة وهو الذي على طرف الذنب, supports a southern-fish tail star, but the transmitted coordinates agree more closely with μ Piscis Austrini, and the identification should be treated as provisional.

7**. Constellation headings.** The constellation names are written in red in the margin of the manuscript (من الدب الاصغر, من التنين, من قيفاوس, من العوّا and so forth) and are transcribed in §10.3; where no rubric stands in the manuscript the cell is left empty rather than supplied editorially.

**8. Astrology.** The final column reflects medieval astrological doctrine. “Favourable” and “unfavourable” are translations of categories transmitted by the source and must not be interpreted as statements of modern astrophysics.

**9. Manuscript collation.** All three manuscript sides containing the 100-star catalogue, fols. 224r–225r, were read in full: the descriptive designations, the marginal constellation rubrics, all numerical columns in abjad, the magnitude qualifiers, the temperament abbreviations and the effects column. The English translation printed in §10.4 is made from that text; Rosenfeld’s Russian edition of 1961 was collated against it entry by entry, and the divergences are listed in §10.4.1 and in the final column of the translation table.

## 11. Discussion

The surviving catalogue offers an unusually compact view of how several strands of medieval astronomical practice converged. It is numerical, descriptive, computational and astrological at once. Its stellar positions belong to mathematical astronomy; the descriptive language preserves an Arabic stellar nomenclature shaped by Greek and indigenous traditions; the magnitude column transmits a Ptolemaic convention; the temperament and effect columns belong to astrology; and its chronological heading embeds the table within the Malikī calendrical framework.

The comparison with Ptolemy is particularly revealing. If 96 longitudes differ from the Ptolemaic values by substantially the same increment, the most economical explanation is not independent observation of all 96 stars but systematic precessional updating. The agreement of 87 stellar latitudes with the corresponding Ptolemaic values reinforces this interpretation. Yet this does not exclude observation altogether. The Malikshāhī astronomical programme was observational in its calendrical work, and individual stellar positions might have been checked, corrected or derived from other sources. The present evidence supports a **hybrid model of inherited data and recalculation**, not a binary opposition between “observation” and “copying”.

This also helps place Khayyām’s work within the broader history of Islamic astronomy. Medieval astronomers inherited the *Almagest* but did not merely preserve it. They recalculated epochs, constructed new zījes, tested parameters and fitted earlier data to new chronological and geographical circumstances. The Malikshāhī catalogue belongs to this tradition of active reuse.

The role of *Dustūr al-munajjimīn* is equally significant. Its incorporation into *Dustūr al-munajjimīn* is the principal reason that this Malikshāhī stellar material is known today. The manuscript thus functions as a repository of otherwise lost astronomical material. Its ten-book organisation, combining calendars, trigonometry, mathematical geography, spherical astronomy, observations, stellar astronomy, astrology and chronology, also demonstrates how later compilers could preserve earlier scientific results outside their original textual environment.

The history of BnF Arabe 5968 adds another dimension. The survival of a scientific work cannot be explained solely by the moment of its composition. This codex passed through users whose interests included jurisprudence, astronomical timekeeping, bibliophilia, manuscript trade and collecting. Jalāl al-Dīn al-Ramlī's position as *muwaqqit* at the Umayyad Mosque provides a particularly striking example of a scientific manuscript remaining relevant to a professional astronomical context centuries after the original Malikshāhī programme.

The Ottoman phase likewise warns against interpreting manuscript movement through geographical names alone. Al-Shīrwānī's nisba originally encouraged an eastward localisation, but Liebrenz's biographical identification situates his collection within Istanbul's elite scholarly and administrative culture. The subsequent gap before Schefer demonstrates how much of a manuscript's history can remain invisible even when major owners are identifiable.

The nineteenth-century transition from Schefer's collection to the Bibliothèque nationale represents a final change in the social life of the codex: a manuscript that had circulated through Islamic scholarly and private libraries entered a European national collection, where modern cataloguing, photography and digitisation eventually made it accessible internationally. The BnF record confirms its Schefer provenance and acquisition in 1899.

The calendrical implications should also be kept conceptually separate from the stellar ones. The Jalālī calendar used the seasonal relationship of the year to the vernal equinox. The fixed-star catalogue, by contrast, deals with stellar positions in an ecliptic coordinate framework. As Bakhromzod has emphasised, the equinoctial origin of the tropical zodiac cannot be identified with the physical constellation Aries because precession continuously changes the relation between the equinox and the background stars [4]. This same precessional phenomenon is precisely what makes the systematic updating of stellar longitudes historically necessary.

Several limitations remain. First, attribution of the catalogue to Khayyām personally cannot be demonstrated entry by entry; it is safer to associate it with the Malikshāhī astronomical project. Secondly, the present English version, although made directly from the Arabic, rests upon a single manuscript witness, and a number of words remain uncertain where the ink has faded or the gutter obscures the text. Thirdly, a limited number of numerical readings and modern stellar identifications remain provisional, notably where the transmitted coordinates conflict with the descriptive identification. Fourthly, direct manuscript collation has clarified the *mizāj* abbreviations and several stellar identifications, but a full star-by-star comparison with Ptolemy and later Islamic catalogues remains desirable. Fourthly, Rosenfeld and Yushkevich's modern stellar identifications deserve rechecking against the Arabic text, Ptolemy and modern historical studies of Arabic star names. Finally, the relationship between this table and the complete lost or unidentified zīj remains only partially reconstructible.

A future critical edition should therefore transcribe BnF Arabe 5968 directly, collate every coordinate against the 1961 edition, identify scribal corrections and marginalia, compare the catalogue systematically with the *Almagest* and later Islamic catalogues, and distinguish the textual layers contributed by the Malikshāhī source from those introduced by the compiler of *Dustūr al-munajjimīn,* including a systematic comparison of the planetary temperaments with Ptolemy's Tetrabiblos, Kūshyār's stellar material, the Zīj-i Īlkhānī, and later zīj traditions.

## 12. Conclusion

The evidence considered here supports several conclusions. No complete copy of the *Zīj-i Malikshāhī* has yet been securely identified; whether the work is entirely lost cannot presently be demonstrated. At the same time, the fragment is substantial enough to illuminate the methods and intellectual environment of the Malikshāhī astronomical programme.

The Jalālī calendar and the stellar catalogue belong to the same broad historical enterprise but served different purposes. The former concerned the organisation of the solar year around the vernal equinox; the latter provided positions and associated information for selected fixed stars. Treating them as identical obscures both the nature of a zīj and the complexity of the Saljuq astronomical project.

The numerical evidence identified by Rosenfeld and Yushkevich—96 longitudes shifted from their Ptolemaic counterparts by about 14°26′ and 87 latitudes agreeing with Ptolemy—strongly suggests systematic adaptation of inherited Ptolemaic coordinates to a later epoch. This practice should be understood as mathematical updating within a living astronomical tradition, not dismissed as passive copying.

Direct comparison with BnF Arabe 5968 also clarifies a previously obscure feature of the catalogue: the *mizāj* column. The manuscript uses abbreviated planetary designations to assign each fixed star the nature of one or two planets. Collation is essential because Rosenfeld's Cyrillic x represents two distinct Arabic letters, خ (Mars) and ه (Venus). Recovering these distinctions provides new evidence for the astrological layer of the Malikshāhī catalogue and permits comparison with other medieval Islamic star tables.

The survival of the fragment also demonstrates the importance of compilation. *Dustūr al-munajjimīn*, composed after the Malikshāhī programme and not written by Khayyām, preserved material that might otherwise have disappeared. The extant BnF manuscript is itself later than the reported 1113 completion of the work; its fourteenth-century palaeographical dating and its ownership history must therefore be distinguished from the chronology of the text it contains.

Finally, the manuscript's route through Syrian scholarly circles, the library of a Damascene *muwaqqit*, Bursa, Ottoman bibliophilic networks, Charles Schefer's collection and the Bibliothèque nationale de France demonstrates that the preservation of medieval science was a process extending across many centuries and institutions. Malikshāhī stellar material survives not through a complete authorial copy but through its incorporation into a later compilation and the subsequent preservation of the codex containing it.

The present transcription, translation and annotated table constitute an intermediate critical step: the Arabic text of all three folios has been read and printed, the English translation is made from that text, and the collation with Rosenfeld's Russian edition both confirms its general accuracy and corrects it at two points, most importantly the longitude of entry 8. A definitive treatment of the Malikshāhī catalogue will require a full critical and diplomatic Arabic edition, independent palaeographical verification of the uncertain readings, and a systematic star-by-star numerical comparison with Ptolemy and other medieval Islamic catalogues.

## Declaration on the Use of Artificial Intelligence

During the preparation of this work, the author used generative-AI tools, including ChatGPT (OpenAI) and Claude (Anthropic), as auxiliary tools for transcription checks, draft translation, comparison of textual readings, exploratory coordinate calculations, and language editing. All readings, numerical values, translations, and conclusions included in the final manuscript were reviewed by the author against the manuscript facsimile and the cited printed sources.

## References

[1] Y. Karamati, “Khayyam, Omar XV. As Astronomer,” *Encyclopaedia Iranica*, published 7 May 2014, updated 29 August 2017.

[2] B. van Dalen, *Islamic Astronomical Tables: Mathematical Analysis and Historical Investigation*. Farnham: Ashgate/Variorum, 2013.

[3] J. Thomann, “The Institution of the Jalālī Calendar in 1079 CE and Its Cohabitation with the Older Persian Calendar,” in S. Stern (ed.), *Calendars in the Making: The Origins of Calendars from the Roman Empire to the Later Middle Ages*. Leiden: Brill, 2021, pp. 210–244. DOI: 10.1163/9789004459694_007.

[4] R. Bakhromzod, “Nowruz, Umar Khayyam, Calendar and Constellations,” *arXiv* 2410.04053 [physics.hist-ph], submitted 5 October 2024. DOI: 10.48550/arXiv.2410.04053.

[5] Bibliothèque nationale de France, Département des Manuscrits, MS Arabe 5968, especially fols. 224r–225r (224a–225a in the a/b foliation), *Dustūr al-munajjimīn*. Digital facsimile, Gallica: https://gallica.bnf.fr/ark:/12148/btv1b525110398/f456.item (images f453–f455 for the catalogue).

[6] Omar Khayyam, *Traktaty*, trans. B. A. Rosenfeld; introductory study and commentary by B. A. Rosenfeld and A. P. Yushkevich. Moscow: Izdatel’stvo vostochnoi literatury, 1961, pp. 225–235 for the Russian translation of the Malikshāhī table and pp. 330–333 for Rosenfeld and Yushkevich’s commentary.

[7] Anonymous, *Dustūr al-Munajjimīn (Canon of the Astronomers)*, facsimile ed. Akbar Irani, with introductory essay by M. Qazvini, English introduction by S. J. Badakhchani, and foreword by Farhad Daftary. Tehran: Mirāth-i Maktūb in association with The Institute of Ismaili Studies, 2019. ISBN 978-6-600-203-180-8.

[8] B. van Dalen, “The Malikī Calendar in the *Dustūr al-munajjimīn*,” in E. Orthmann and P. G. Schmidl (eds.), *Science in the City of Fortune: The Dustūr al-munajjimīn and Its World*. Berlin: EB-Verlag, 2017, pp. 117–135.

[9] E. Orthmann and P. G. Schmidl (eds.), *Science in the City of Fortune: The Dustūr al-munajjimīn and Its World*. Bonner Islamstudien 39. Berlin: EB-Verlag, 2017.

[10] B. Liebrenz, “The History and Provenance of the Unique *Dustūr al-munağğimīn* Manuscript, BnF Arabe 5968: A Re-assessment,” *Journal of Islamic Manuscripts*, vol. 11, 2020, pp. 28–42. DOI: 10.1163/1878464X-01101002.

[11] Bibliothèque nationale de France/Biblissima, catalogue record for MS Arabe 5968: fourteenth-century naskh manuscript, 346 folios, formerly in the collection of Charles Schefer and acquired in 1899.

[12] Ptolemy, *Ptolemy’s Almagest*, trans. and annotated by G. J. Toomer. London: Duckworth, 1984, Books VII–VIII.

[13] J. Evans, “On the Origin of the Ptolemaic Star Catalogue,” *Journal for the History of Astronomy*, vol. 18, 1987, pp. 155–172.

[14] P. Kunitzsch and Y. T. Langermann, “A Star Table from Medieval Yemen,” *Centaurus*, vol. 45, nos. 1–4, 2003, pp. 159–174. DOI: 10.1111/j.1600-0498.2003.450115.x.